# Deliverable

**Grant Agreement number:** 611742

**Project acronym:** PASTEUR4OA

**Project title:** Open Access Policy Alignment STrategies for European Union Research

**Funding Scheme:** FP7 – CAPACITIES – Science in Society

**Project co-ordinator Organisation:** **EKT/NHRF**

**E-mail:** **tsoukala@ekt.gr**

**Project website address:** www.pasteur4oa.eu

| | |
|---|---|
| **Deliverable No.** | 3.1 |
| **Deliverable Name** | Report on policy recording exercise, including policy typology and effectiveness and list of further policymaker targets |
| **Lead Beneficiary** | EOS |
| **Dissemination Level** | PU |
| **Due Date** | M12 |

# D 3.1 – Report on policy recording exercise, including policy typology and effectiveness and list of further policymaker targets

PASTEUR4OA is an FP7 project funded by the EUROPEAN COMMISSION

# Working Together to Promote Open Access Policy Alignment in Europe

## Work Package 3 report: Open Access Policies

March 2015

PASTEUR4OA Project

OPEN ACCESS POLICY:
NUMBERS, ANALYSIS, EFFECTIVENESS

Alma Swan, Yassine Gargouri, Megan Hunt and Stevan Harnad
Enabling Open Scholarship

# Table of Contents

# Figures and tables

**Figures**

**Tables**

## 1. Executive summary

The PASTEUR4OA project is focused on Open Access policy developments and is undertaking a number of activities relating to policy, including mapping policies and policy-related activities, and engaging with policymakers and providing them with information about the general policy picture and what makes a policy effective.

Work Package 3 involved a set of tasks as follows:

- Describe and enumerate the policy picture in Europe and around the world
- Rebuild ROARMAP, the registry of OA policies, including the development of a new, detailed classification scheme that describes policy elements
- Collect data on the levels of Open Access material in institutional repositories around the world
- Measure policy outcomes and analyse what elements of a policy contribute to its effectiveness

The project sought out policies that exist but had not been registered in ROARMAP, and added more than 250 new entries to the database. The total number of policies globally is now 663 (March 2015), 60% of them from Europe. Of these, approximately two-thirds are institutional policies and about 10% are funder policies. Over half are mandatory, *requiring* some action rather than simply requesting it and over 60% of these mandatory policies are European.

ROARMAP, the policy registry, has been rebuilt with a new classification scheme for policies that records far more detail about them than before and permits much more extensive search functionality than previously. The scheme includes criteria for deposit and licensing conditions, rights holding, embargo lengths and 'Gold' Open Access publishing options. Links to policy documents are provided. Repository managers at policy institutions were contacted to check that we had the correct details for their policy and where necessary corrections were made. As it stands, at the end of this period of concentrated and meticulous work, ROARMAP reflects an accurate and detailed picture of the Open Access policy situation around the world.

The project also examined policy effectiveness. Three main exercises were undertaken.

First, deposit rates were measured for articles in the repositories of both mandated and non-mandated institutions, and compared to the total number of articles published from these institutions. The material was identified as Metadata-Only, Full-Text, Open Access and Restricted Access. Open Access and Restricted Access are subsets of Full-Text and together comprise the whole of that category. Restricted Access means full-text articles that are showing only their metadata, with the text itself closed off, and are usually in this state for a period of embargo.

Across all institutions, more than three-quarters of published articles are not deposited at all, 8% are Metadata-Only, 3% Restricted Access and 12% Open Access. The rates vary by discipline. Deposit of Open Access material was over four times as high (14%) for institutions with a mandatory policy than for those without (3%). The top 20 institutions (all mandated) in terms of amount of repository content are listed. The top five are the University of Liège (Belgium), Instituto Politecnico de Bragança (Portugal), the National Institute of Oceanography (India), University of Pretoria (South Africa) and the University of Minho (Portugal).

Second, the time lag between publication and deposit of articles (deposit latency, which may be negative if the article is deposited before publication) was measured. Open Access items tend to be deposited later than Restricted Access ones, and latency periods tend to be longer in mandated institutions than in non-mandated ones (though deposits themselves are four times higher), probably because authors who deposit voluntarily are self-motivated and will do it early.

Third, we examined the deposit rate in relation to different policy criteria:

- Positive correlations were found between Open Access and Restricted Access deposit rates and the following policy criteria: ***Must deposit, Cannot waive deposit, Link to research evaluation, Cannot waive rights retention, Must make item Open Access***
- Negative correlation was found with ***Cannot waive Open Access***
- Significant correlation was found between Open Access deposit rate and ***Must deposit*** and ***Cannot waive deposit***

Fourth, we examined the correlation between deposit latency (specifically, the latency of deposit within the first year after publication) and different policy criteria. There is positive correlation between early deposit and ***Mandate age, Cannot waive rights retention*** and ***deposit*** immediately. We found significant correlation between early Open Access deposits and the age of the mandate: that is, the longer a mandatory policy has been in place, the more effective it can become.

As the numbers stand at the moment (March 2015), there are not yet enough OA policies to test whether other policy conditions would further contribute to mandate effectiveness. The current findings, however, already suggest that it would be useful for future mandates to adopt these conditions so as to maximise the growth of OA.

This analysis provides a list of criteria around which we recommend policies should align:

- Must deposit (i.e. deposit is mandatory)
- Deposit cannot be waived
- Link deposit with research evaluation

## 2. Introduction and background to the work in Work Package 3

Open Access policies first began to appear in 2002 in the form of a sub-institutional policy from the School of Electronics & Computer Science at the University of Southampton, UK. Since then they have been growing in number and currently there is a total of over 600, some of them in second or third iterations. At the beginning of the PASTEUR4OA project we knew there were hundreds of policies, but we were not completely confident that we knew exactly how many, so one of the key pieces of work for the project has been to rectify this. We have undertaken a wide scale search for policies that were not registered in ROARMAP, the database of Open Access policies, and have registered them so that the database is as comprehensive as possible.

Initially, policies were made almost exclusively in institutions, but once the first research funder policy, a mandatory policy from the Wellcome Trust, appeared in 2005, others began to follow. For example, the National Institutes of Health in the USA adopted a policy in 2005: it was not a mandatory one at the time of its inception but was strengthened a couple of years later to a mandate when it became clear that it in its original form it was ineffective in delivering Open Access[1]. Policies from other research funders followed quickly behind in ensuing years, along with increasing numbers from research institutions and parts thereof.

At the start of the PASTEUR4OA project, we knew that there had been considerable policy development and adoption by European institutions and funders but we were not sure of the exact numbers or the level of growth. Nor did we know how the European picture compared with the rest of the world. Are European institutions and funders leading the way, or lagging behind? This was one thing we set out to determine.

Policies vary hugely. Some are mandatory, at least with respect to one or more specific actions in the provision of Open Access. Conversely, many policies do not mandate Open Access in any respect, but instead simply *encourage* or *request* certain behaviours of authors. Different policies may even have different aims, resulting in differing wording and emphasis. We wanted to understand the nature of existing policies, and to ascertain whether some sort of typology could be developed to establish some clarity across the diversity of these documents. Consequently, we created a comprehensive classification scheme and undertook a detailed analysis of the policies in ROARMAP, classifying each of them according to the criteria in the scheme.

Europe, of course, now has the Horizon 2020 (H2020) Open Access policy, a very significant development for Open Access in this region since it applies to an €80 billion research funding programme and to research projects and programmes across all ERA (European Research Area) countries.

1 The NIH policy has, along with that of the Wellcome Trust, been strengthened yet again: both funders have introduced compliance monitoring and have taken further steps to remind grant-holders of their obligations under the policy.

The main elements of the policy are:

- Open Access is mandatory for peer-reviewed publications
- The policy is a **'Green' OA** mandate (repositories)
  - Publish as normal in subscription-based journals
  - Place author's copy in OA repository
- For **'Gold'** OA, the policy permits payments from grants for OA journal publication fees where they are levied
- The policy says nothing about OA for **monographs**, but there may be some attention to this issue as time goes on
- The policy is very definite about Open Research Data, announcing an **Open Data pilot** for the H2020 programme

The European Commission has recommended[2] that Member States follow its example and make OA policy where they have not already done so, and that these policies should emulate the H2020 one. The overall thrust of the PASTEUR4OA project is to stimulate policy development in line with this Commission recommendation, work that involves engaging and informing policymakers, providing an evidence base for designing policy and making arguments for aligned policy development.

It is important to understand the needs for, and benefits of, aligned policies. Many researchers obtain funding for their research programme from more than one funder, sometimes several. If each of these has, in essence, a policy that is the same as the others – that is, requires researchers to do a certain set of things in the same way – then he researchers will have a clear set of requirements with which to comply. If the policies differ significantly – that is, they each require different things of the researchers – then the researchers perceive a degree of complication and are less well-disposed towards compliance. If the policies differ very significantly from each other then compliance will reduce still further. It is also important that the OA policies should be aligned on the specific policy conditions that make the policy effective in terms of compliance rate and timing.

Some Member States do already have national funder policies, but their conditions differ in detail. The project has looked at these to try to answer the key questions – are they very different, essentially the same, quite similar, or similar in at least the critical ways? Might they be aligned? There are many examples of policies changing and evolving over time: what are the chances of existing policies that are not aligned with the H2020 policy becoming more like it?

Importantly, how could we find out and document all this? The existing policy registry service, ROARMAP, listed around 375 policies as the PASTEUR4OA project started, but it was known to be incomplete. Some entries were also known to be inaccurate, for example, as a result of the database not being updated when a policy was changed. One objective of the PASTEUR4OA project, then, was to

2 European Commission Recommendation of 17.7.2012 on access to and preservation of scientific information: http://ec.europa.eu/research/science-society/document_library/pdf_06/recommendation-access-and-preservation-scientific-information_en.pdf

revamp ROARMAP to provide an accurate and up-to-date reference service with vastly enhanced information about each policy.

We then needed to understand what makes some policies effective and others not. Detailed guidelines for policymaking have already been produced by the Harvard Open Access Project[3]. There have been a couple of empirical studies already carried out to look at this. We knew from the study by Gargouri *et al*[4] several years ago that mandatory policies work better than ones that simply encourage authors to comply: the evidence from that study showed that the average deposit rate for institutions without an OA mandate is 15% while that for institutions with a mandate is 60%.

A follow-up study by the same team[5], using data from the Spanish policy effectiveness database, MELIBEA, explored further and found that here things correlated positively with policy effectiveness as measured by *deposit rate* and *deposit latency* (the length of time that elapses between publication in a journal and deposit of the item in a repository). The policy conditions that correlate positively with effectiveness measured in these terms are (i) immediate deposit required (ii) deposit required for performance evaluation and (iii) unconditional opt-out allowed for the OA requirement but no opt-out for deposit requirement.

The PASTEUR4OA project's policy classification for ROARMAP provided a rich database about policies that could be used to carry out an analysis of policy effectiveness. To do this we needed to know the levels of Open Access material that institutions are succeeding in collecting, which the project measured. Then we analysed which elements of policy correlate with success in delivering high levels of OA content. We now understand much better the elements of a policy that are critical to its effectiveness and this evidence can be used to show best practice in policymaking. The findings are reported in Section 5.

This document reports on the work carried out by the PASTEUR4OA project, which consisted of several main tasks as follows:

- Describe and enumerate the policy picture in Europe and around the world
- Rebuild ROARMAP, the registry of OA policies, including the development of a new, detailed classification scheme that describes policy elements
- Collect data on the levels of Open Access material in institutional repositories around the world
- Measure policy outcomes and analyse what elements of a policy contribute to its effectiveness

---

3 Harvard Open Access Project: Good practices for university open-access policies: http://cyber.law.harvard.edu/hoap/Good_practices_for_university_open-access_policies

4 Gargouri, Y., Hajjem, C., Lariviere, V., Gingras, Y., Brody, T., Carr, L. and Harnad, S. 2010. Self-Selected or Mandated, Open Access Increases Citation Impact for Higher Quality Research. *PLOS ONE*, 5 (10). e13636. http://www.plosone.org/article/info:doi/10.1371/journal.pone.0013636 http://eprints.ecs.soton.ac.uk/18493/

5 Vincent-Lamarre P, Boivin J, Gargouri Y, Lariviere V and Harnad S (2014) Estimating Open Access Mandate Effectiveness: I. The MELIBEA Score (submitted for publication). http://arxiv.org/abs/1410.2926

# 3. ROARMAP: the registry of Open Access policies

## 3.1 The history of ROARMAP

The ROARMAP database[6] was set up in 2004 and hosted alongside stablemates ROAR (the Registry of Open Access Repositories), Citebase and the OpCit project at the School of Electronics & Computer Science at the University of Southampton in the United Kingdom. At the time the acronym stood for Registry of Open Access Repositories Material Archiving Policies, a database of 'Green' (self-archiving) Open Access policies implemented by institutions and research funders.

The database recorded the numbers and growth of policies and provided both basic details of and links to the policies, and visualisations (it could create graphs on the fly) of the growth of policy types. The types recorded were: institutional policies, funder policies, multi-institutional policies, sub-institutional (school or department-level) policies and policies from organisations that are both funders and research-performing bodies. This original ROARMAP allowed browsing and searching by type, country, continent, and the date policies were adopted.

Registration of policies was carried out by Southampton personnel or by policymakers themselves. Links to the policy document were encouraged, along with brief descriptions of policies. Some entries provided neither, however.

By mid-2014, ROARMAP contained approximately 375 policies with around 95 from funders, 225 from institutions, 50 from sub-institutional units (departments, schools or faculties) and a handful from multi-institutional consortia. By this time the acronym stood for Registry of Open Access Repositories Mandatory Archiving Policies because the aim at the time was to record the growth of Open Access *mandates* from institutions and funders – that is, policies where Open Access is a mandatory requirement.

At this point in mid-2014, the PASTEUR4OA project began work to update and upgrade ROARMAP. In brief, a search was carried out worldwide for policies not yet recorded in the database; a new and detailed classification scheme was developed and existing and newly-found policies were categorised accordingly; and the technical specification was enhanced to give the database much greater functionality.

At the time of writing, ROARMAP contains some 663 policies (research-performing organisations 461, sub-institutional policies 69, research funders 72, joint funder/research organisations 53, multiple-institutional policies 8). Each policy has been classified according to the new scheme developed by the PASTEUR4OA project.

A description of the activities carried out by the project to revamp ROARMAP is given below. The final point to be made here is that, given that a number of the policies recorded in the ROARMAP database

6 http://roarmap.eprints.org/

are not mandatory in nature, the acronym now stands for *Registry of Open Access Repository Mandates and Archiving Policies*.

## 3.2 The new iteration of ROARMAP

Work carried out by the PASTEUR4OA project to revamp ROARMAP has been the following:
- A search to find policies that were not registered in ROARMAP and to add them, thus increasing its comprehensiveness
- The development of a new, detailed classification scheme for policies
- The classification of all policies in the database according to the new scheme
- The re-launch of the new ROARMAP with increased content and greater functionality

### *3.2.1 Ensuring comprehensiveness*

Project partners searched for policies that were in existence but not registered in ROARMAP. They did this by searching the web, by making enquiries directly with possible policyholders and by using their contacts. Each partner covered a different world region so the exercise had a global reach.

Through this exercise, well over 100 policies were discovered that had not been registered in ROARMAP. Of these, 23 are mandatory (see below for more on the nature of policies). The holders of these policies were consulted for permission to add their policy to ROARMAP or encouraged to add their policy themselves.

In addition, around 90 unregistered policies were discovered through the consultation with repository managers (see Section 3.2.3 below) and were added to ROARMAP, bringing the total number of policies to 663 at the time of writing (March 2015), an increase of some 288 entries in ROARMAP as a result of the project's work. The database is now far more comprehensive than before the PASTEUR4OA project began.

### *3.2.2 The new policy classification scheme*

The objective of this exercise was to reassess the information held in ROARMAP and update the data, where appropriate, to reflect the changing landscape of Open Access policy and to provide an accurate database containing details of policies from institutions around the globe. To do this, we had to check each policy and record its characteristics.

For this, we developed a new, detailed classification scheme for policies. As well as general information about each policy, such as date of adoption, the policymaker type, and organisational data, the new scheme covers criteria for deposit and licensing conditions, rights holding, embargo lengths and 'Gold' Open Access publishing options. The full set of criteria is given in Appendix 1.

Wherever a policy was available it was examined and classified to determine the scope of the policy in terms of deposit conditions, strength of mandate (if applicable), and publishing specifications. This task became a lengthy and time-consuming process as not all policies were available or easily accessible in the first instance. Project partners also noted comments on their research process for finding each policy, as well as comments on the nature, availability and specific wording of each policy document.

Some lessons were learned from this, and these are summarised in Section 3.2.4.

### *3.2.3 Classification of the policies in ROARMAP*

Records were exported from the original ROARMAP database into a working spreadsheet and the policies of each institution classified according to the new scheme (see Appendix 1), which can be broadly divided into five categories:

- Institutional particulars
- Dates relating to policy implementation
- Criteria for deposit and licensing conditions
- Rights holding
- Embargo lengths and publishing options

Approximately 70% of the policy documents for the policies already classified in ROARMAP were found through links from ROARMAP, web searches, blogs and direct communication with institutions and were able to be examined, while the remaining 30% were either in the draft stage, planned for the future or the institution in question had no policy at the present time.

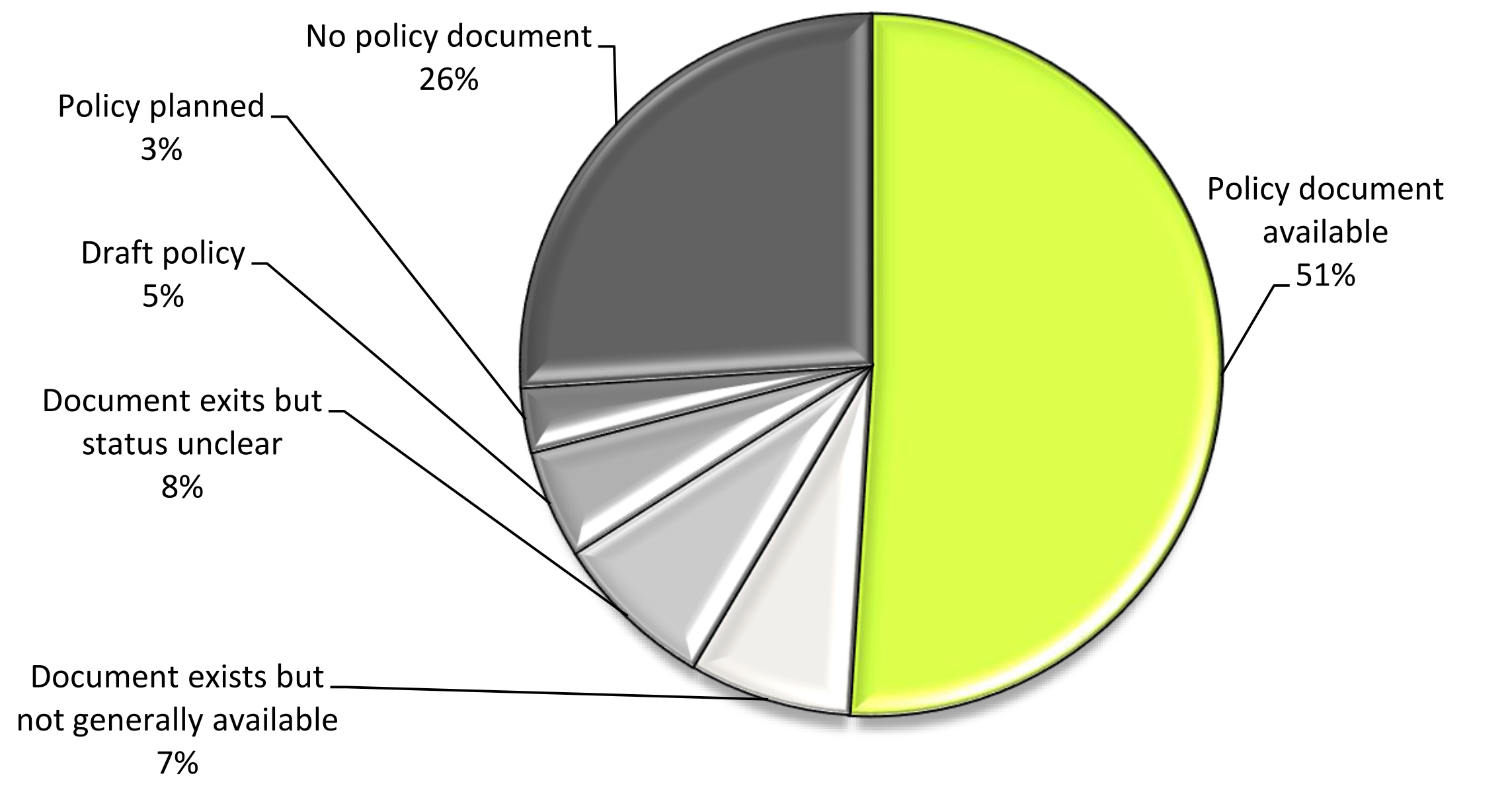

***Figure 1: Availability of policy documents for policies in the original ROARMAP database***

Following initial classification and examination of the data, logical judgements were applied to the resulting information before it could be re-entered into the new version of the database. The development of a logic structure included the removal of contradictory statements – for example, ensuring a policy stating that deposit was not mandatory was not also classified as including a deposit waiver. During this process a number of these inaccuracies were corrected and any obviously omitted information identified or corrected where possible.

Where there were still concerns regarding the accuracy of information, contact details of repository managers at each institution were obtained and they were then contacted by email and asked to verify the status of their Open Access policy by classifying their policy according to the new scheme. Their responses endorsed the accuracy of the classification work by PASTEUR4OA project workers, but also provided much needed clarification in areas of ambiguity as well as more precise details – for example, implementation dates and URLs.

### *3.2.4 Issues identified*

During the classification work a number of issues were identified which posed a challenge to the project, mainly related to the lack of an available policy document.

In the first instance, where a policy was not readily discoverable through an institution's webpages or repository, details were sought using a web search or by attempting to contact the repository manager or institution directly, with mixed results.

A high proportion of the problematic records in the original ROARMAP database contained dead links, and in some cases when further research was carried out no other links to policy could be found. In a small number of instances policies were only accessible through institutional intranets and only to those affiliated with the organisation.

On rare occasions the institutional website or repository was under construction and links had been moved and deleted. This also included institutions which had merged or changed names since the original version of ROARMAP was established. In a number of instances information regarding policy had to be gleaned from blog posts, online press releases or news items included on institutional webpages or within the general open access community.

Direct correspondence with repository managers was undertaken (as mentioned above) and the policy details verified through them were added to the new version of the database. This also proved to be an effective strategy where policy documentation was lacking in details or vaguely worded regarding the strength of mandate or conditions of deposit and provided much needed clarification in some areas. It was also an extremely useful exercise when classifying policies that are only available in local languages and it helped to reduce the need for translation of a number of policies.

Approximately 5% of policies within ROARMAP are in draft format, and therefore the database will require a degree of on-going curation to ensure it remains up-to-date.

# 4. The Open Access policy picture in Europe and globally

## 4.1 The number of Open Access policies

The current global picture with respect to the number of policies is shown in Table 1 and Figure 2 below. Note that these numbers are for all policies, not just mandatory ones. Europe leads in terms of policies in relation to research intensity (measured by articles published). Europe has approximately 25% of the world's researchers (FTEs) with North America (USA and Canada) in second place (22%)[7], but Europe has more than 2.5 times the number of OA policies as North America.

| Region | Policies |
|---|---|
| Europe | 389 |
| North America | 145 |
| Central and South America | 34 |
| Africa | 16 |
| Asia | 40 |
| Oceania | 39 |

***Table 1: Number of Open Access policies worldwide***

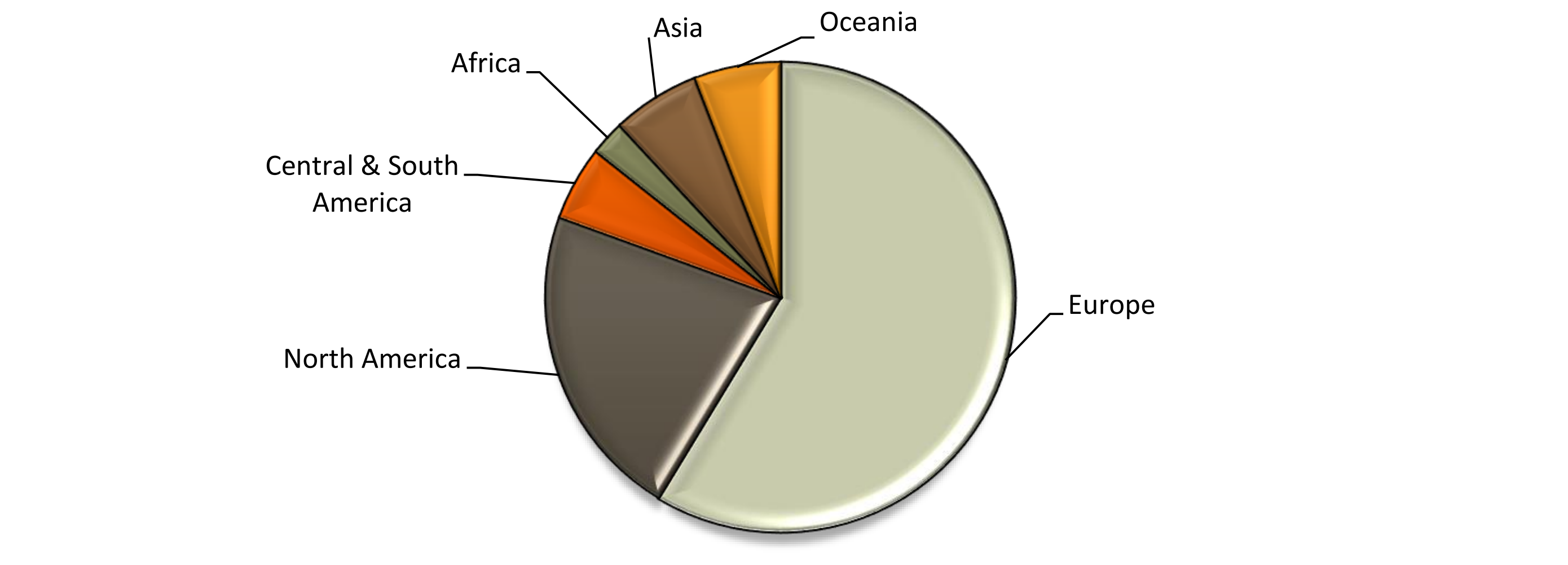

***Figure 2: Number of Open Access policies worldwide***

About two-thirds of the total policies cover whole-institutions, with further institutional policies covering sub-units of institutions and institutional consortia. In addition, there are policies from funders. The whole picture is shown in Table 2 and Figure 3 below.

7 OECD Main Science & Technology Indicators: http://www.oecd.org/science/msti.htm

| Policymaker type | Policies |
|---|---|
| Research funder | 72 |
| Research institution | 461 |
| Research funder and institution | 53 |
| Multiple research institutions (consortia) | 8 |
| Sub-unit of research institution | 69 |
| **Total** | **663** |

***Table 2: Open Access policies worldwide by policymaker type***

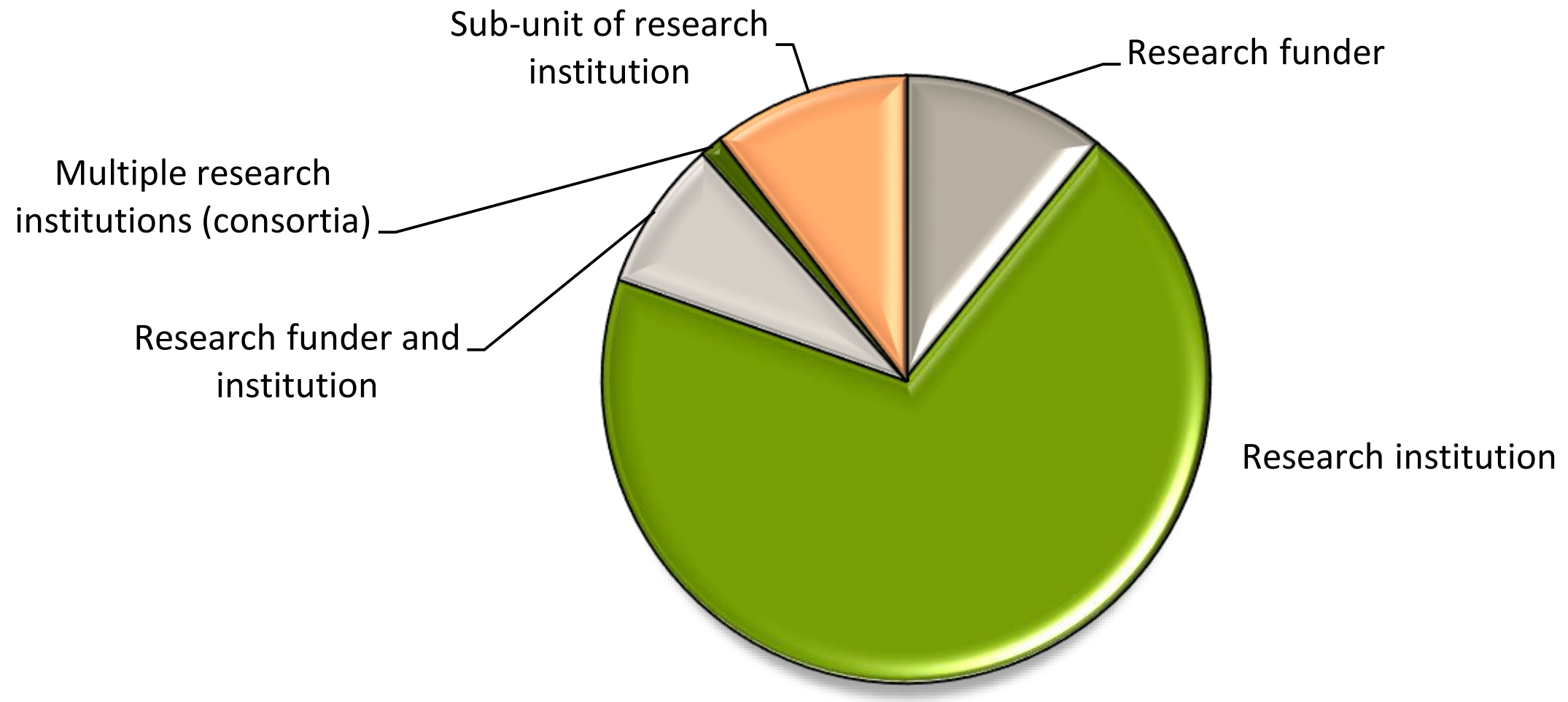

***Figure 3: Open Access policies worldwide by policymaker type***

## 4.2 Policy criteria (conditions)

Many policies are not mandatory but simply encourage or request authors to provide Open Access for their outputs. Approximately half of existing policies are mandatory in that they *require* a particular behaviour. In almost all cases this is articulated in the form *'authors must deposit'* when referring to repository-mediated Open Access (Green OA), often with an encouragement to publish their articles in journals (Gold OA). In a small number of cases, policies have some form of words implying that authors must publish their work in Open Access form in journals (Gold OA) wherever possible (with Green OA being the fall-back alternative in this case).

| Criterion (Green OA) | Number of policies | Criterion (Gold OA) | Number of policies |
|---|---|---|---|
| Deposit in repository required (Green OA) | 381 | OA publishing required | 2 |
| Deposit in repository requested | 140 | Recommended alternative to Green OA | 97 |
| Deposit in repository not specified | 141 | Permitted alternative to Green OA | 101 |
| | | Not specified/other | 463 |
| **Total** | **663** | | **663** |

***Table 3: Open Access policies: Green and Gold OA criteria***

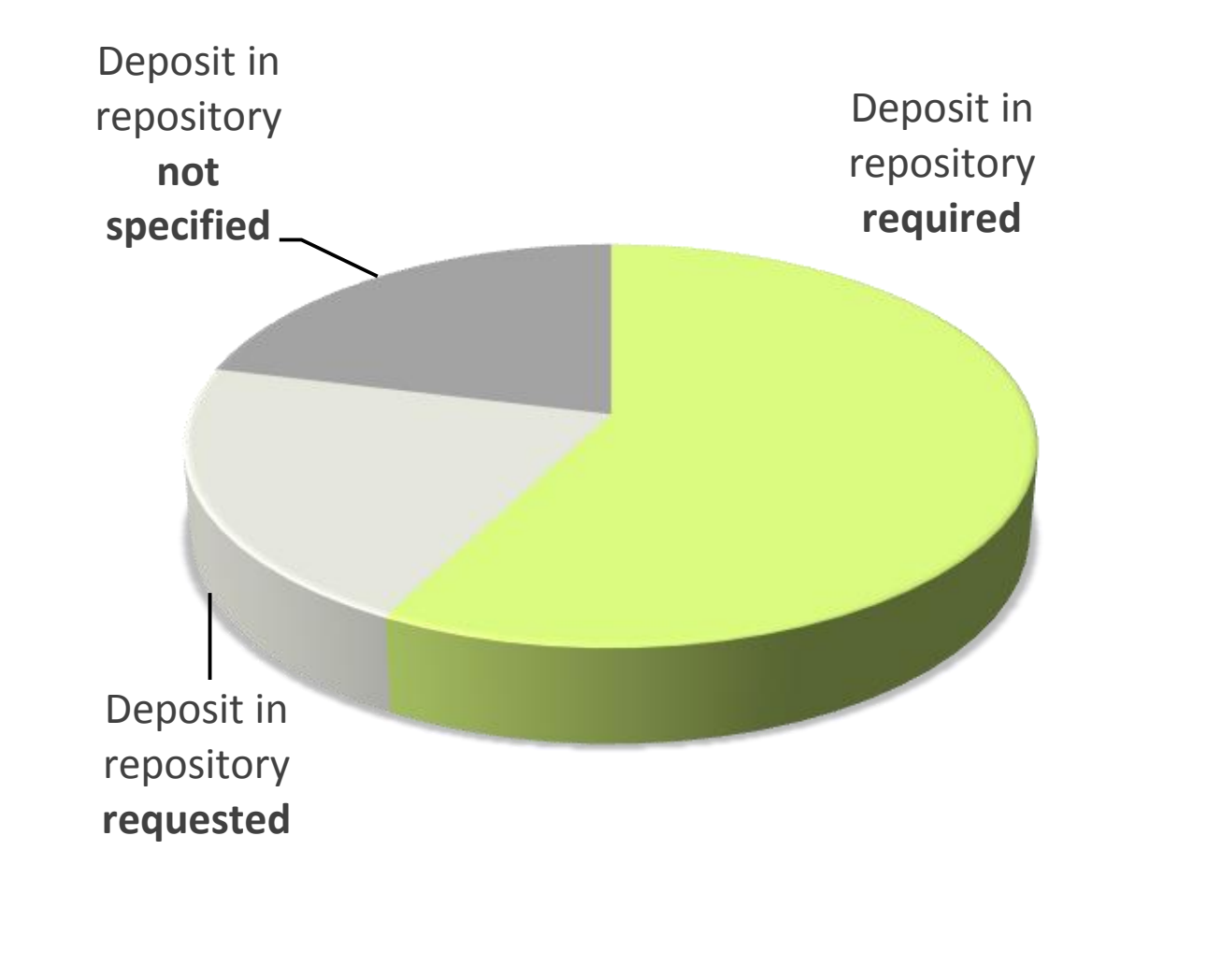

*Figure 4: Green OA policy requirements*

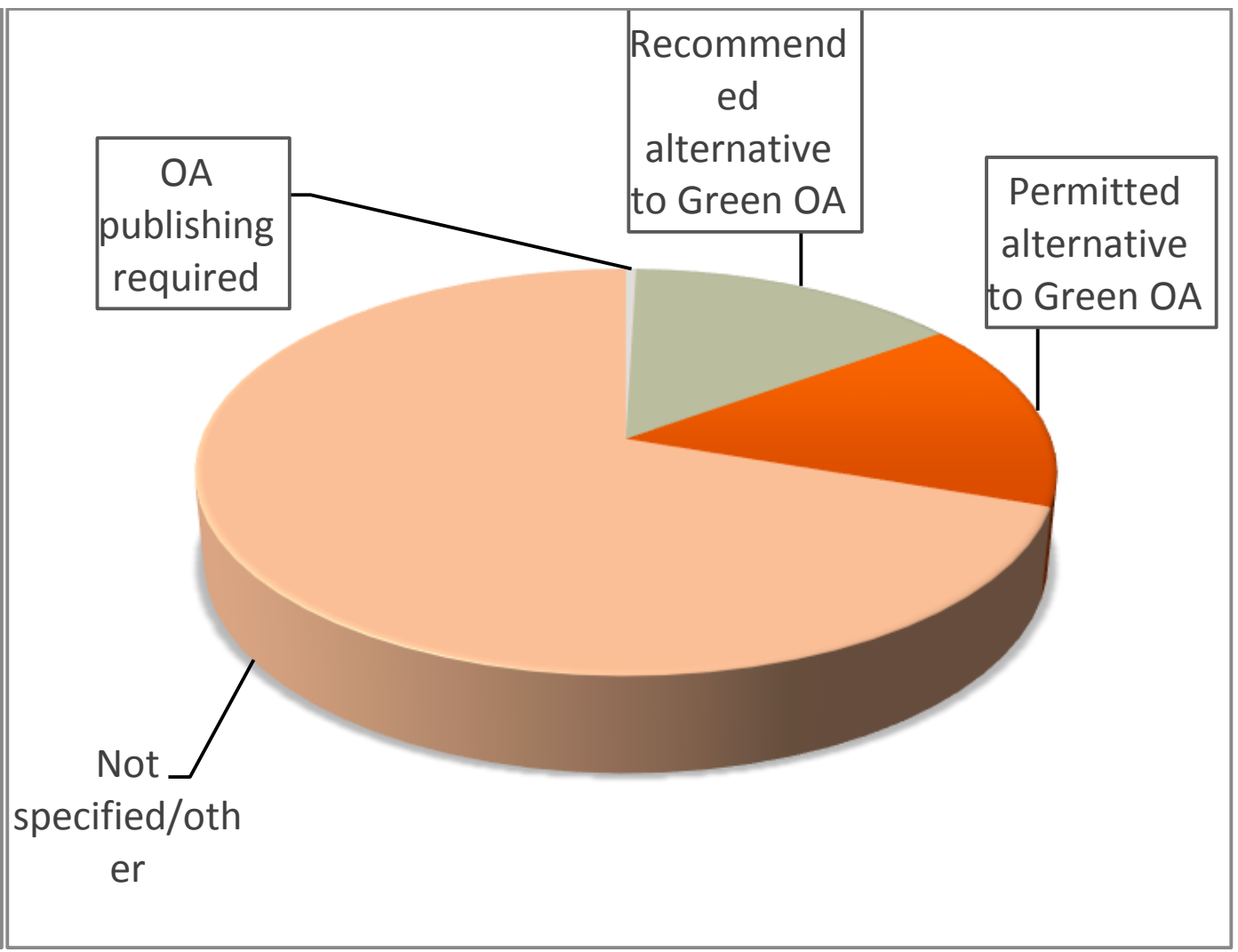

*Figure 5: Gold OA policy requirements*

The situation for research funders is given in Table 4 and Figures 6 and 7. Note that the numbers for policies from funder/institution combinations are not included here.

| Criterion (Green OA) | Number of policies | Criterion (Gold OA) | Number of policies |
|---|---|---|---|
| Deposit in repository required (Green OA) | 49 | OA publishing required | 1 |
| Deposit in repository requested | 12 | Recommended alternative to Green OA | 18 |
| Deposit in repository not specified | 11 | Permitted alternative to Green OA | 11 |
| | | Not specified/other | 43 |

*Table 4: Open Access policies: Green and Gold OA criteria – research funders*

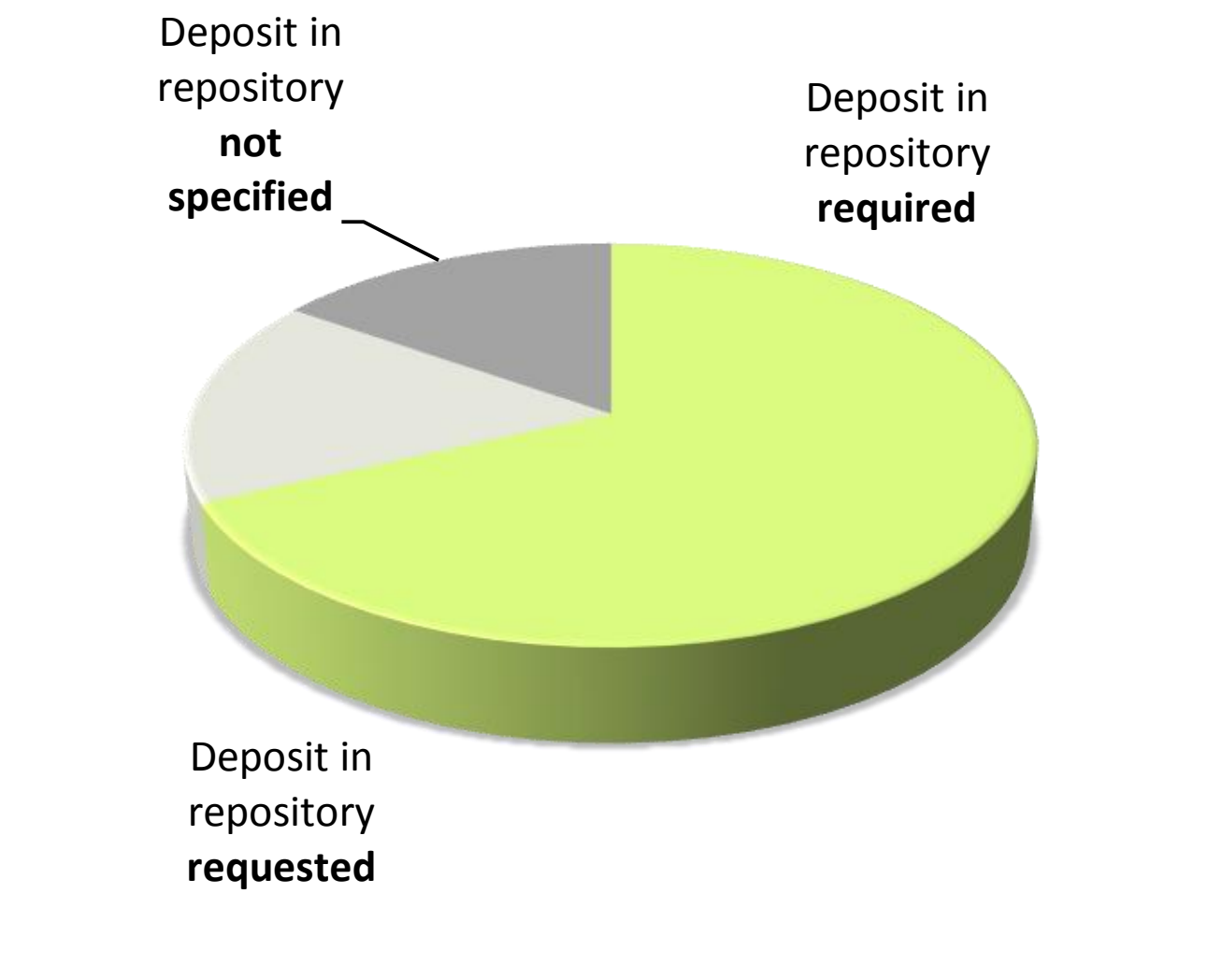

*Figure 6: Green OA policy requirements: research funders*

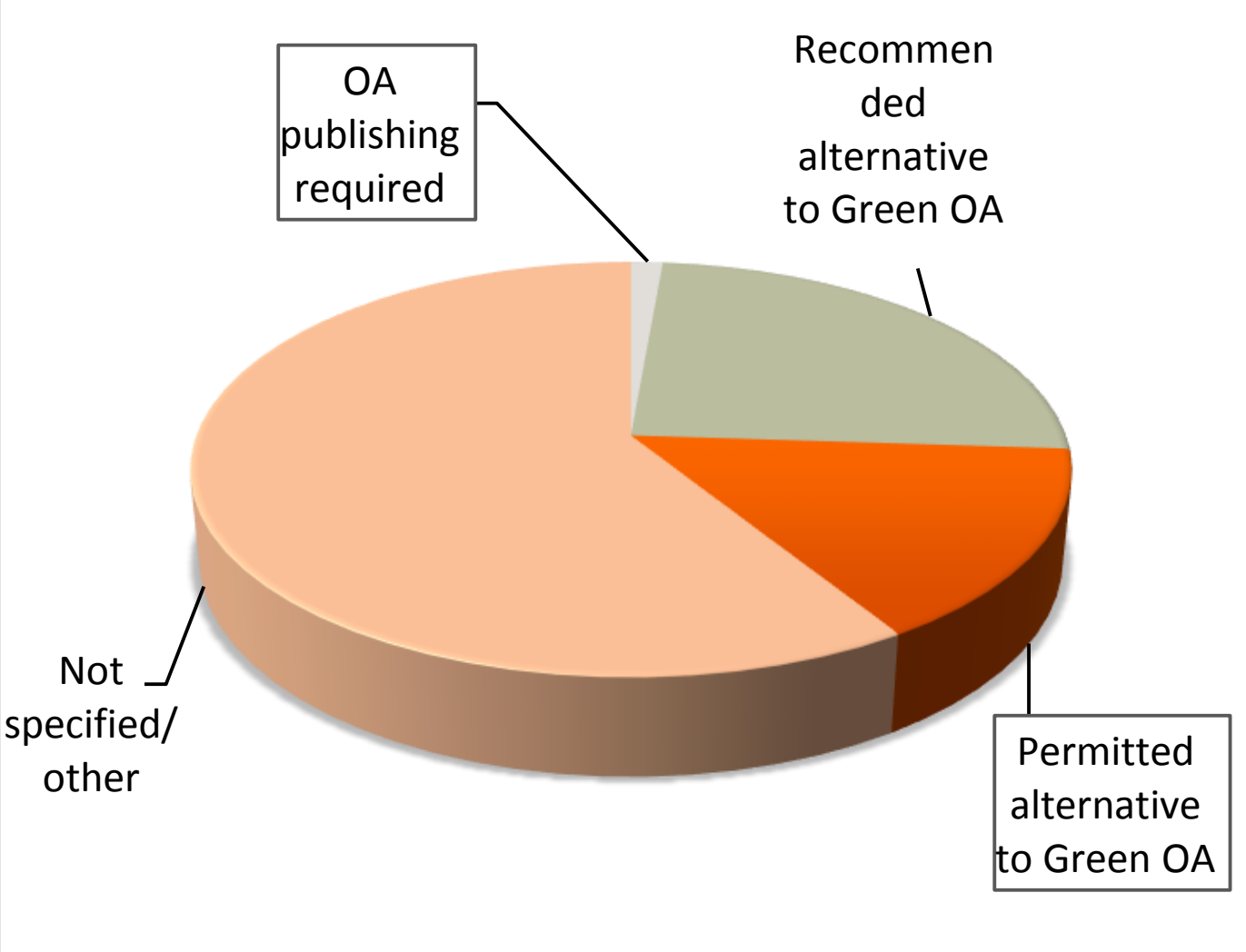

*Figure 7: Gold OA policy requirements: research funders*

The situation for research institutions (whole institutions plus multiple institutions and sub-units of institutions) is given in Table 5 and Figures 8 and 9. Note that the numbers for policies from funder/institution combinations are not included here.

| Criterion (Green OA) | Number of policies | Criterion (Gold OA) | Number of policies |
|---|---|---|---|
| Deposit in repository required (Green OA) | 310 | OA publishing required | 2 |
| Deposit in repository requested | 124 | Recommended alternative to Green OA | 75 |
| Deposit in repository not specified | 104 | Permitted alternative to Green OA | 82 |
| | | Not specified/other | 379 |

***Table 5: Open Access policies: Green and Gold OA criteria – research institutions***

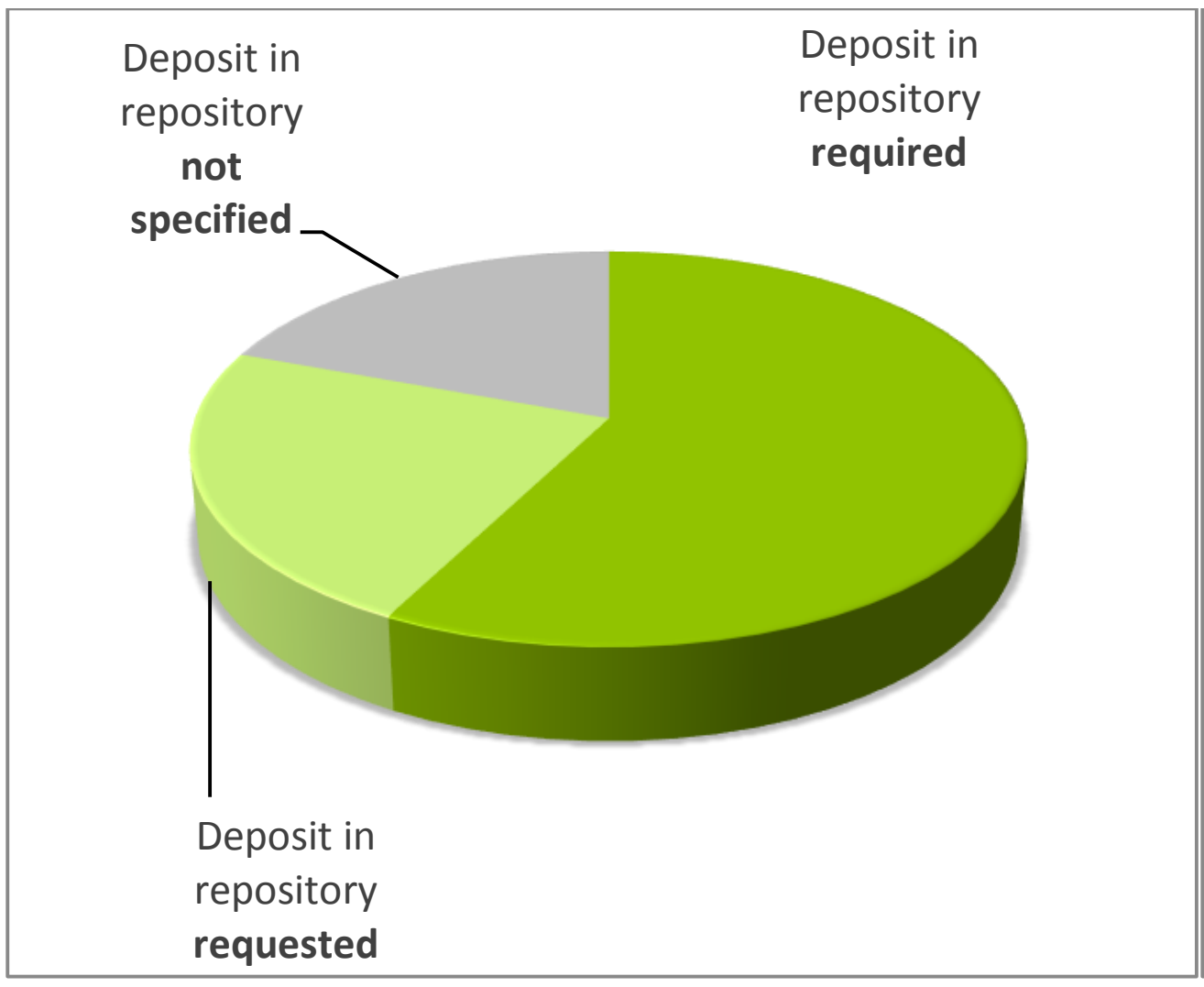

***Figure 8: Green OA policy requirements: institutions***

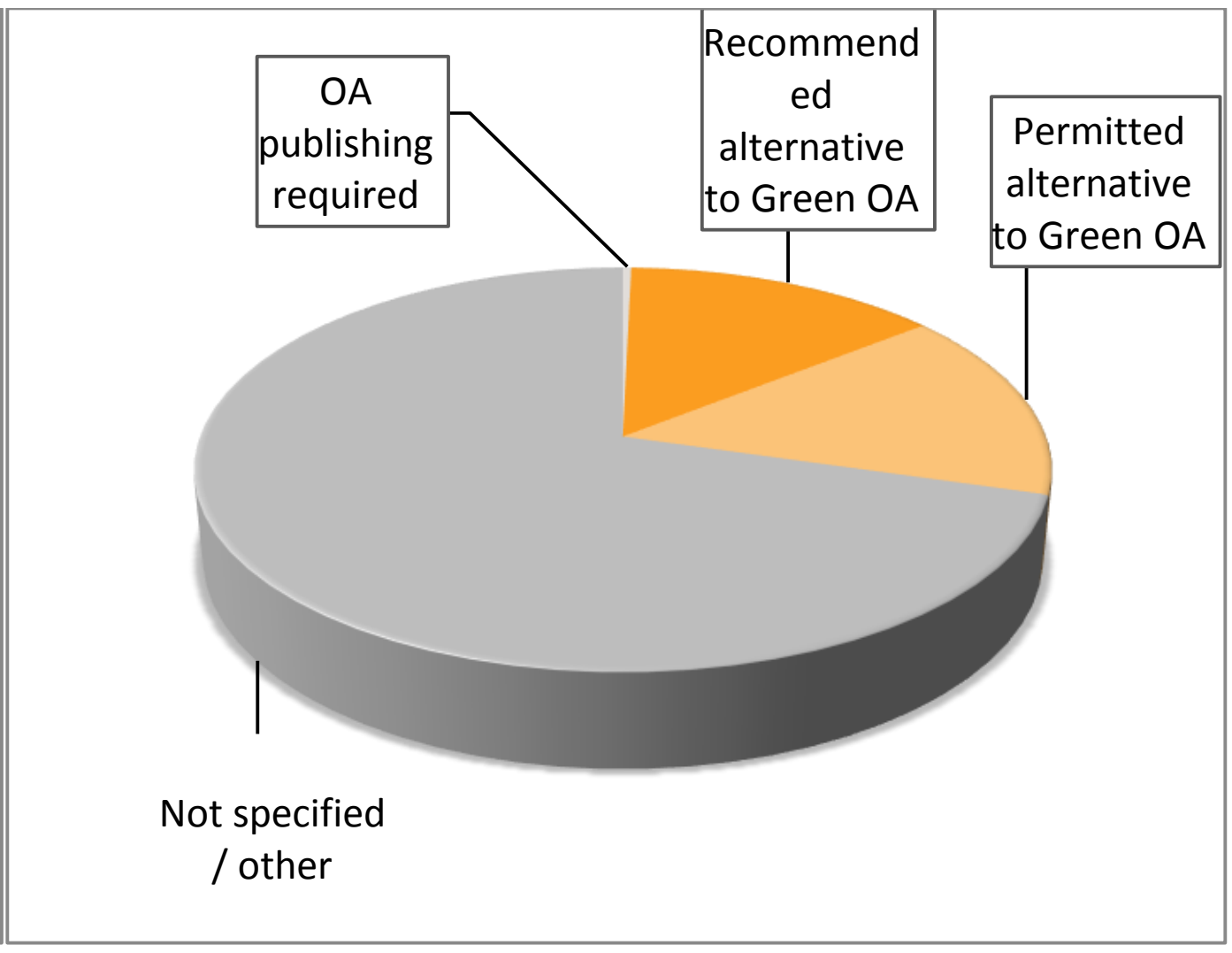

***Figure 9: Gold OA policy requirements: institutions***

## 4.3 Mandatory Open Access policies

It was known from earlier studies (see Section 2) that mandatory policies work much better than voluntary ones. The situation with respect to mandates worldwide is shown in Table 6 and Figure 10 below. A mandate is defined for this purpose as a policy that requires deposit of articles in a repository (Green OA) or requires Open Access publishing of articles (Gold OA).

| Geographical location | Mandatory policies |
|---|---|
| Africa | 10 |
| Asia | 24 |
| Central and South America | 18 |
| Europe | 237 |
| North America | 75 |
| Oceania | 20 |

***Table 6: Mandatory Open Access policies worldwide by geographical region***

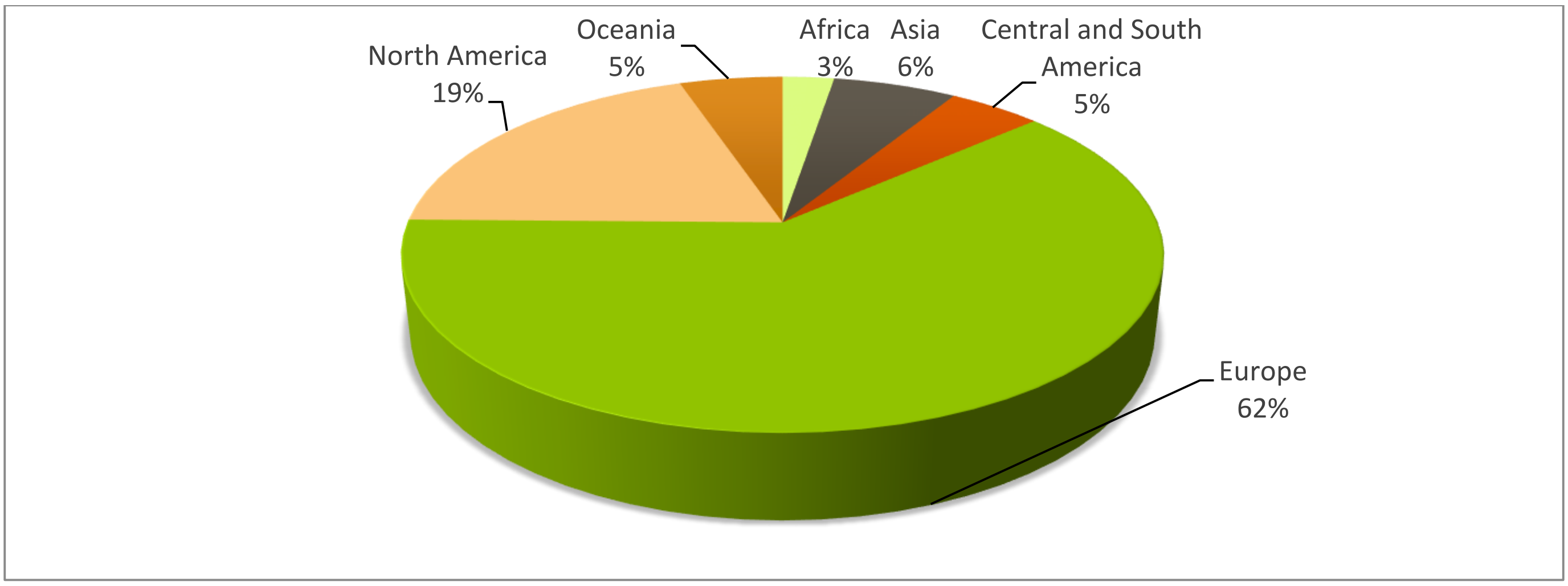

***Figure 10: Mandatory Open Access policies worldwide by geographical region***

## 4.4 Time of deposit

Policies vary with respect to the time-point they specify for deposit of items. The charts below show the numbers specifying different time-points.

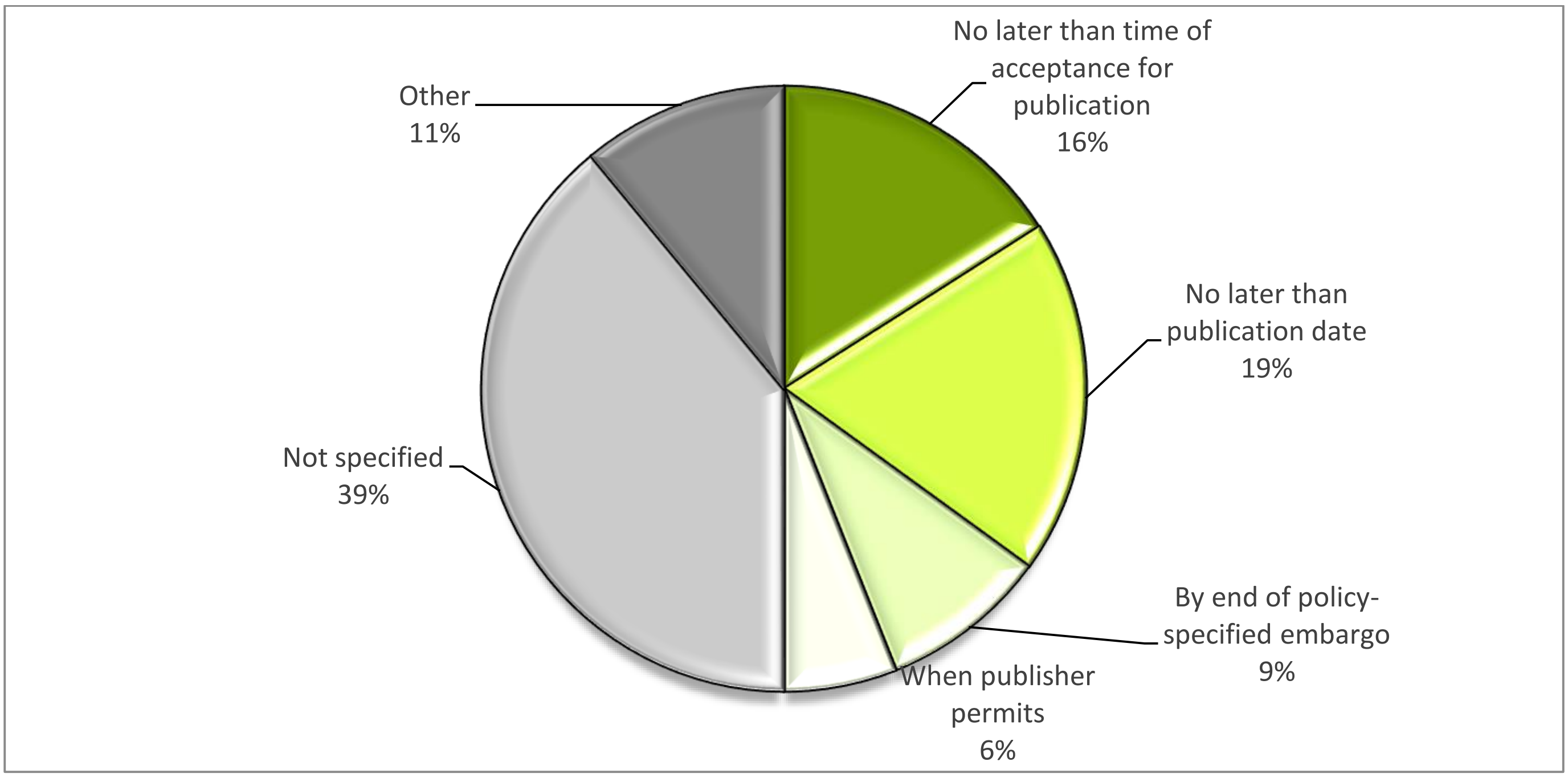

***Figure 11: Time-point for deposit specified by mandatory policies***

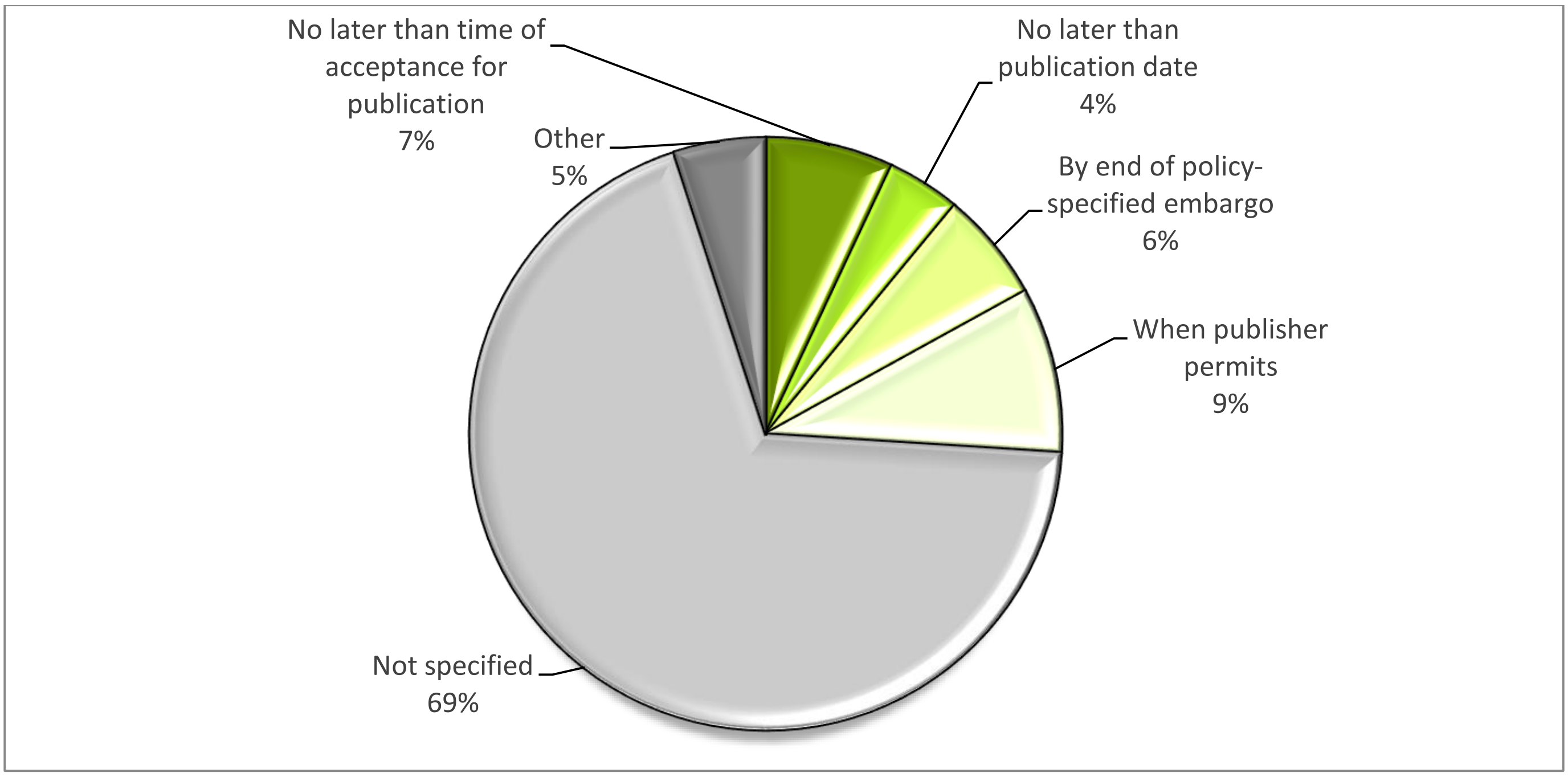

*Figure 12: Time-point for deposit specified by policies that 'request' rather than mandate deposit*

# 5. Policy effectiveness

## 5.1 Introduction

Policy is made on Open Access because researchers have not spontaneously made their work openly available, despite the proven benefits to themselves of doing so. The Web is now more than twenty years old, yet estimates of the level of Open Access continue to indicate that we are far from having an open research literature. In 2009, Björk and colleagues found the percentage of OA to be just over 20%[8]; in 2013, Archambault estimated it to be around 50%[9], but in 2014 Chen came up with a figure of 38%[10] and Khabsa & Giles 24%[11] (all these figures are Web-wide levels, whereas this present study measured the levels in institutional repositories).

It is worth noting that none of these studies took into account the timing of access – that is, whether the papers were openly available from or near the time of publication or whether they had been made open months or years later. The latest data from Björk *et al* show that, given that 62% of journals permit immediate self-archiving by their authors, 4% impose an embargo of 6 months and 13% an embargo of 12 months, almost 80% of articles could already be openly available within a year of publication. But they are not: the Open Access corpus rather stubbornly remains the lesser part of the literature.

Hence the policy thrust. We have reported earlier in this document on the numbers and growth of OA policies, both from institutions and research funders. The numbers have grown from that very first sub-institutional mandate in 2002[12] to over 600 today, and yet the proportion of Open Access material does not reflect that. Many policies are apparently ineffective in delivering Open Access, whereas some are very effective indeed.

The PASTEUR4OA project sought to understand this better. What policy types successfully deliver Open Access? What clauses in a policy are the most effective in this regard? The exercise was carried out on institutional policies only, because research funder policies are difficult to monitor due to the dearth (hitherto) of metadata of a quality that enables articles from particular funder programmes or projects to be tracked accurately.

The study carried out five activities for the period 2011-2013:

8 Björk B.-C., Welling P., Laakso M., Majlender P., Hedlund T., *et al*. (2010). Open Access to the scientific journal literature: Situation 2009. *PLoS ONE 5(6):* e11273. doi:10.1371/journal.pone.0011273
http://www.plosone.org/article/info:doi/10.1371/journal.pone.0011273
9 Archambault, E., Amyot D., Deschamps P., Nicol A., Rebout L., and Roberge G. (2013) Proportion of Open Access peer-reviewed papers at the European and world levels – 2004-2011. Report for the European Commission by Science Metrix.
http://www.science-metrix.com/pdf/SM_EC_OA_Availability_2004-2011.pdf
10 Chen X (2014) Open Access in 2013: reaching the 50% milestone. *Serials Review* **40(1)**, 21-27.
11 Khabsa M. and Giles C.L. (2014) The number of scholarly documents on the public Web. *PLoS ONE*, **9(5)** e93949 doi:10.1371/journal.pone.0093949
http://journals.plos.org/plosone/article?id=10.1371/journal.pone.0093949
12 School of Electronics & Computer Science, University of Southampton, UK

- Measured the contents of institutional repositories for institutions with a mandatory OA policy ('mandated institutions') and a comparison set of institutions that do not have a mandatory policy ('non-mandated institutions')
- Measured the amount of repository content that is Full Text (FT), Open Access (OA: Open Access, full-text items), and Restricted Access (RA: ie embargoed full-text items), by institution, discipline and year
- Ranked mandated institutions according to Open Access deposit level
- Ascertained the deposit delay (deposit latency: the time between publication date and deposit), by institution and by discipline
- Determined the correlation (by multiple regression analyses) between individual policy conditions and deposit rate (ie percentage of published output deposited as Open Access or Restricted Access) to test which policy criteria (independent variables) correlate with deposit percentages and deposit latency (dependent variables).

The methodology used for this work is given in Appendix 2.

## 5.2 Institutional repository deposit rates

### *5.2.1 Overall deposit rates*

The average Full Text (FT) deposit rate in institutional repositories for 2011-2013 across all WoK-indexed journal articles for the institutions in this study (mandates and non-mandated) was 15.5%. This 15.5% was comprised of 12.4% as Open Access material and 3.1% as Restricted Access (RA) material.

More than three-quarters (76.4%) of articles from institutions are not deposited at all, and a further 8% are Metadata-Only (MO) deposits.

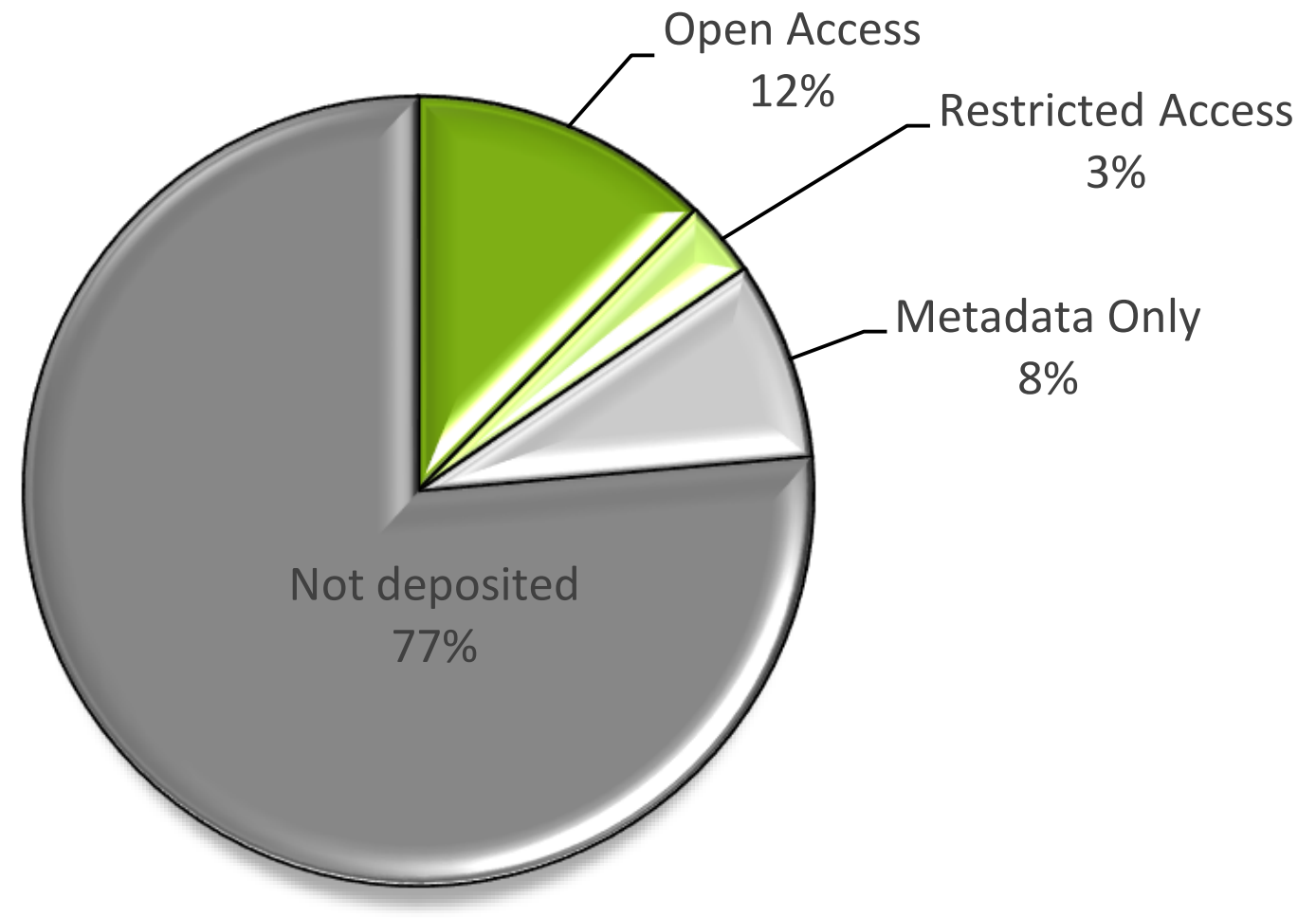

***Figure 13: Institutional repository content types: repository average (across mandated and non-mandated institutions)***

### 5.2.2 Deposit rates by discipline

The percentage of published articles deposited in institutional repositories varies by discipline. Data are presented in tabular and graphical form below.

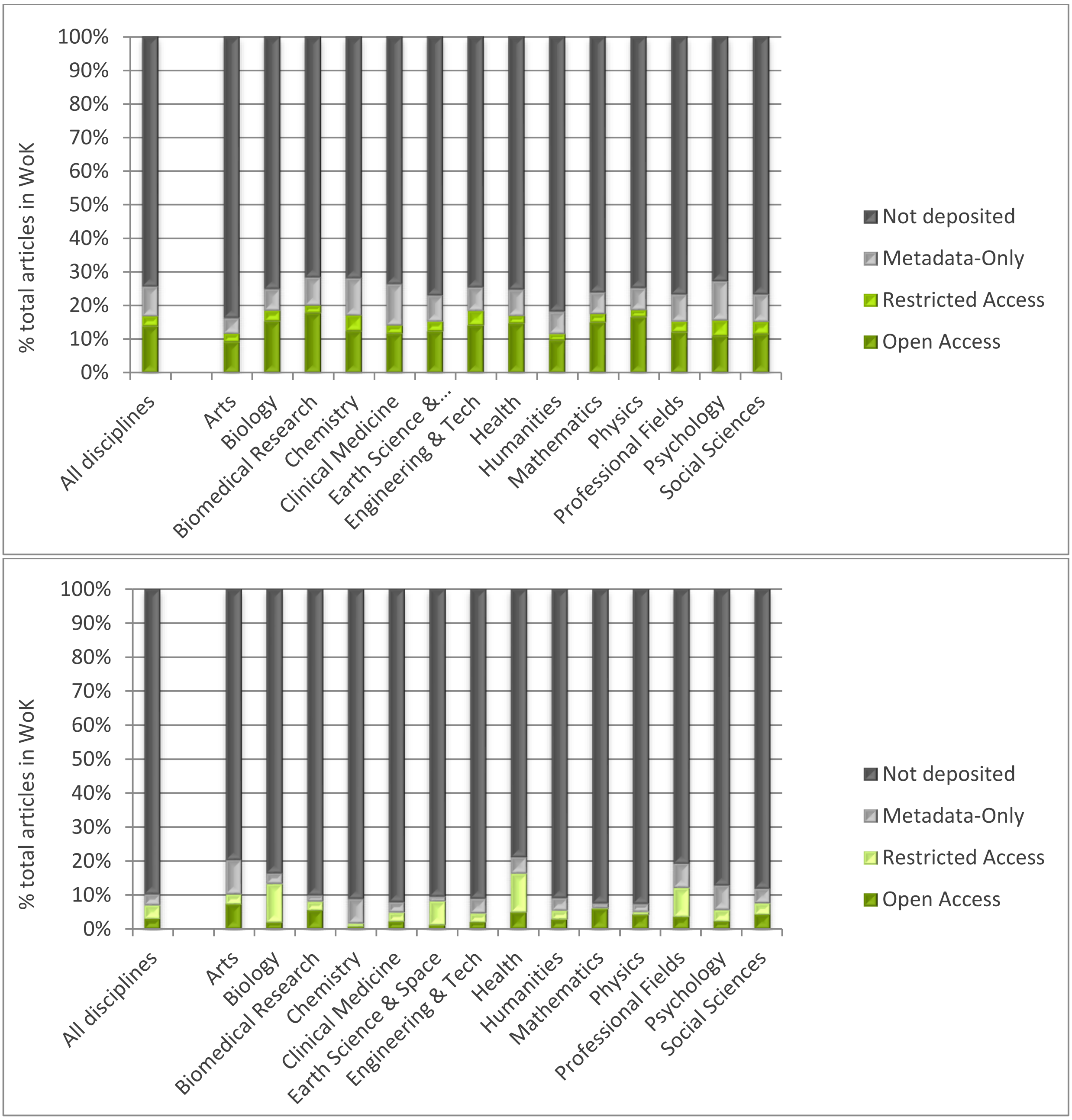

***Figure 14: Institutional repository deposit levels for mandated (top) and non-mandated (bottom) institutions, by discipline***

Deposit percentages in non-mandated institutions are much lower, as expected in the light of previous studies. There are very successful subject repositories in some fields, notably mathematics/physics and biomedicine, and some content that is deposited by authors in those, either spontaneously by authors or as a result of funder mandates that specify subject repositories, may be lost to institutional repositories as a result. This should be taken into account when interpreting the data in Figure 14. Of the 295 institutional mandates, 284 specify that deposit must be made in the institution's repository, 9 allow deposit in any suitable repository and 2 do not specify a locus of deposit.

### *5.2.3 Mandated and non-mandated institutions: per cent deposit*

The OA deposit rate (percentage) was over four times as high (13.8%) for mandated institutions as for non-mandated ones (3.0%). The Restricted Access deposit rate was about the same whether mandated or not (3% and 4% respectively). Mandated institutions also had a higher rate of deposit of Metadata-Only material (8.8%) than non-mandated institutions (3.3%). In the table below, Full-Texts are the sum of Open Access and Restricted Access items (i.e. FT = OA + RA). Metadata-Only items are a separate category, as are the Not Deposited items.

| | Mandated institutions % total outputs in WoK | Non-mandated institutions % total outputs in WoK |
|---|---|---|
| Full-Text deposits | 16.8 | 7.0 |
| Open Access deposits | 13.8 | 3.0 |
| Restricted Access deposits | 3.0 | 4.0 |
| Metadata-Only deposits | 8.8 | 3.3 |
| Not deposited | 74.3 | 90.0 |

*Table 7: Content of mandated and non-mandated institutional repositories (IRs)*

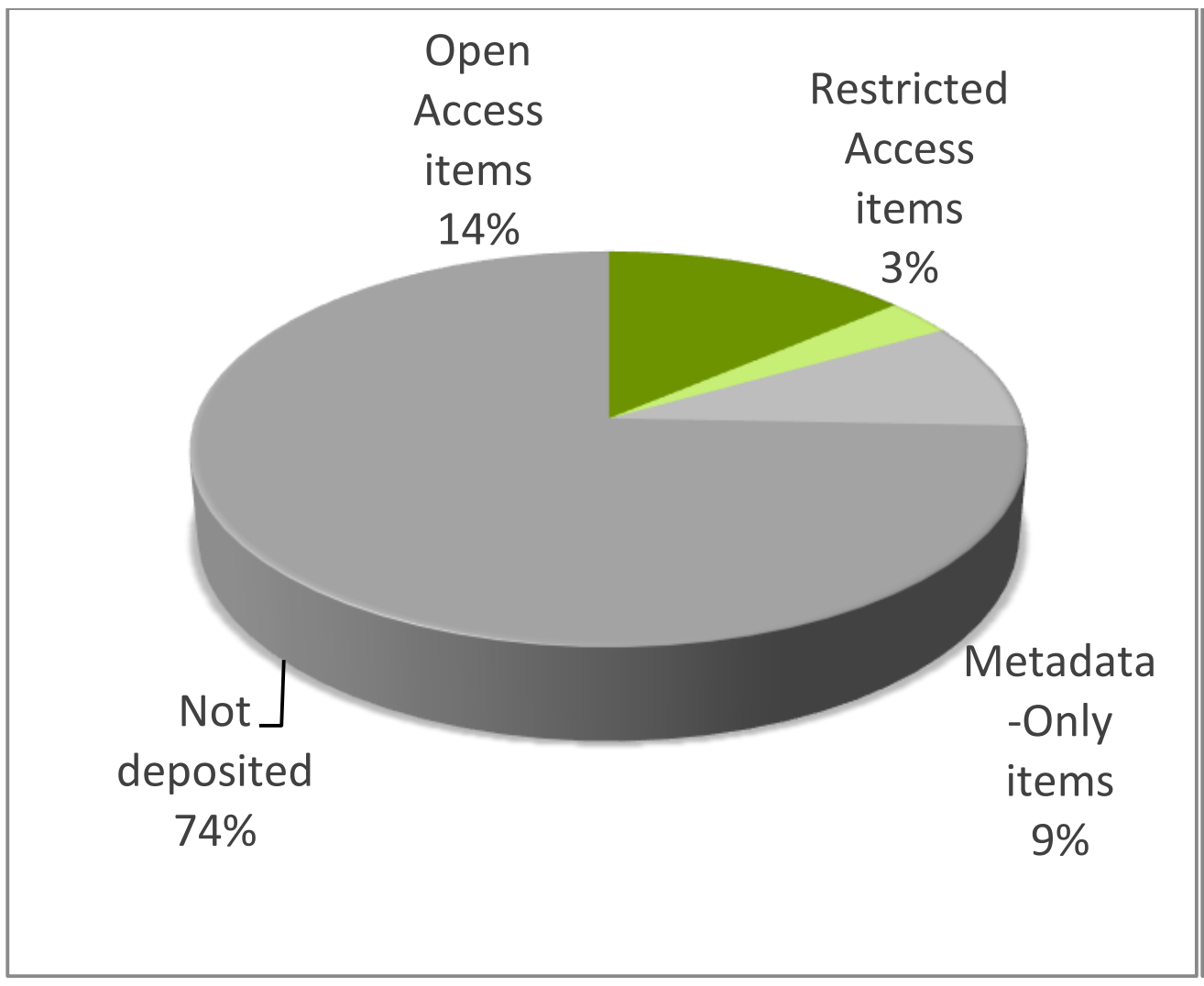

*Figure 15: Mandated IR deposits*

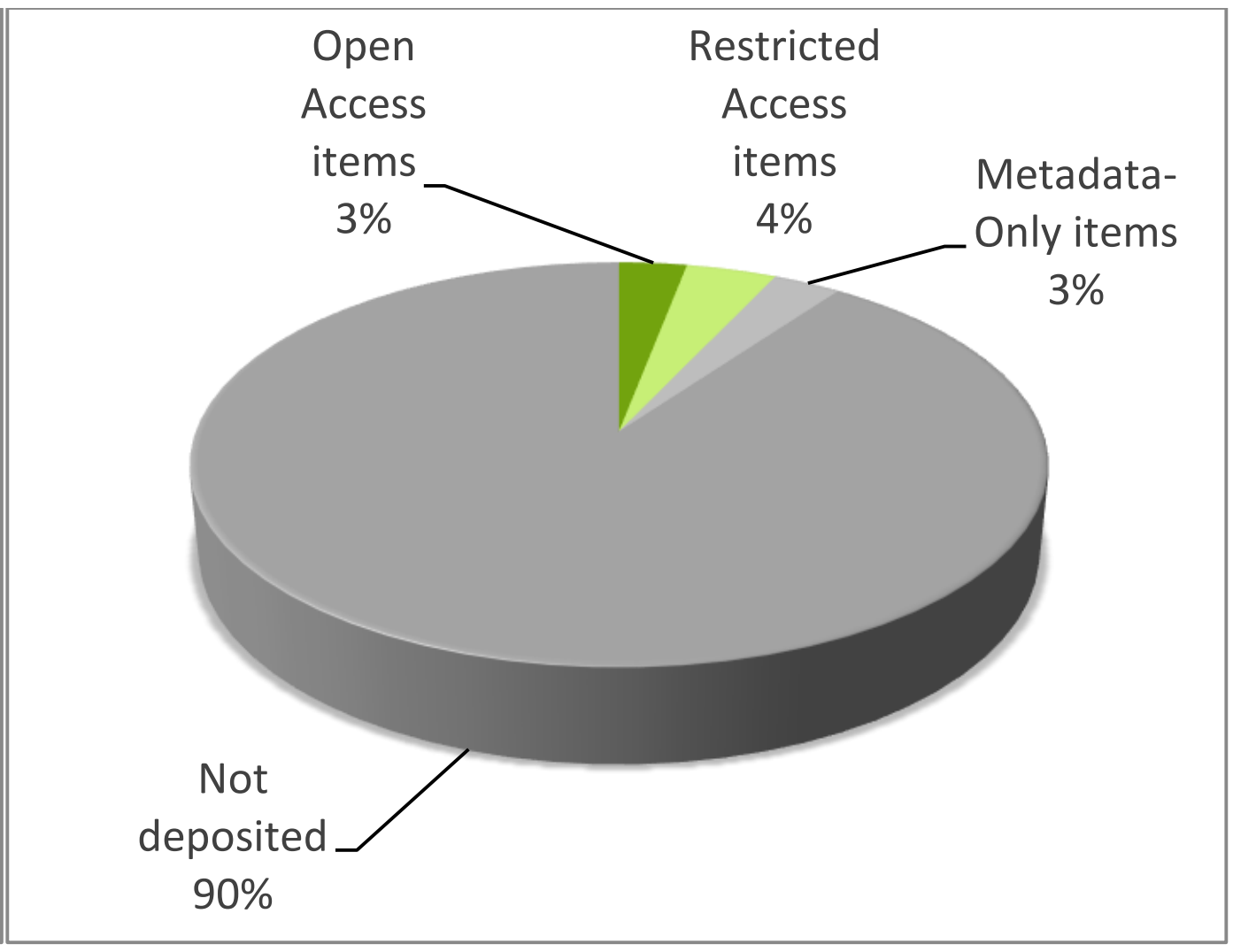

*Figure 16: Non-mandated IR deposits*

### *5.2.4 Deposit rates in individual mandated institutions*

Table 8 shows the percentages of Full-Text (Open Access plus Restricted Access), Open Access, Restricted Access and Metadata-Only deposits in the repositories of the institutions that have at least 50 articles indexed in WoK in the years 2011-2013. They are rank-ordered by *per cent Full-Text.*

| Institution | Country | Number of articles in WoK | Total FT (OA + RA) % | Open Access % | Restricted Access % | Metadata-Only % | Not deposited % |
|---|---|---|---|---|---|---|---|
| **All institutions** | | **70,642** | **43.0** | **35.8** | **7.2** | **5.2** | **51.8** |
| University of Liege | Belgium | 4,240 | 87.0 | 37.0 | 50.0 | 0.1 | 12.9 |
| Instituto Politecnico de Braganca | Portugal | 267 | 85.8 | 56.9 | 28.8 | 0.0 | 14.2 |
| National Institute of Oceanography | India | 462 | 79.7 | 79.7 | 0.0 | 0.2 | 20.1 |
| Universidade do Minho | Portugal | 3,021 | 62.3 | 39.1 | 23.2 | 0.0 | 37.7 |
| University of Pretoria | South Africa | 3,335 | 60.4 | 60.4 | 0.0 | 0.0 | 39.6 |
| University of Nairobi | Kenya | 655 | 60.0 | 60.0 | 0.0 | 6.4 | 33.6 |
| Queen Margaret University, Edinburgh | United Kingdom | 150 | 57.3 | 14.7 | 42.7 | 8.0 | 34.7 |
| University of Luxembourg | Luxembourg | 761 | 55.8 | 18.9 | 36.9 | 0.5 | 43.6 |
| Queensland University of Technology | Australia | 3,558 | 49.1 | 44.4 | 4.7 | 35.0 | 15.9 |
| Belgorod State University | Russia | 189 | 45.0 | 45.0 | 0.0 | 0.0 | 55.0 |
| University of Stirling | United Kingdom | 1,301 | 41.7 | 15.7 | 26.1 | 0.0 | 58.3 |
| Universidade de São Paulo - USP | Brazil | 21,080 | 41.1 | 41.1 | 0.0 | 3.4 | 55.5 |
| Universitat Politècnica de Catalunya | Spain | 3,394 | 39.8 | 13.9 | 25.9 | 0.6 | 59.6 |
| University of Surrey | United Kingdom | 2,613 | 35.6 | 29.4 | 6.2 | 0.1 | 64.3 |
| Massachussetts Institute of Technology (MIT) | United States | 14,019 | 32.3 | 32.3 | 0.0 | 0.0 | 67.7 |
| University of Salford | United Kingdom | 822 | 31.1 | 7.5 | 23.6 | 5.2 | 63.6 |
| University of Loughborough | United Kingdom | 2,615 | 30.7 | 30.7 | 0.0 | 0.0 | 69.3 |
| Brunel University | United Kingdom | 2,244 | 30.1 | 30.1 | 0.0 | 0.1 | 69.8 |
| Universidad Politécnica de Madrid | Spain | 3,051 | 29.6 | 26.2 | 3.4 | 0.0 | 70.4 |
| University of Bath | United Kingdom | 2,847 | 28.5 | 28.0 | 0.5 | 55.4 | 16.1 |

***Table 8: Content of institutional repositories of individual mandated institutions with more than 50 articles indexed in WoK in the period 2011-2013***

## 5.3 Deposit latency

The deposit latency is the length of the time gap between the time of publication and that of deposit, the latter being recorded in the repository metadata. A positive latency indicates that the article was deposited after the publication date: a negative latency indicates that the article was deposited earlier than the publication date. Deposit latency can be calculated for Open Access and Restricted Access items.

### *5.3.1 Deposit latency by discipline*

Deposit behaviour might be expected to vary between disciplines and certainly publisher embargo lengths play a role here. We found that the average deposit latency for Open Access deposits in the humanities was 6.8 months, whereas in clinical medicine it was 14.1 months. For Restricted Access deposits the values vary from 2.9 months in the humanities to 8.7 months in biology. Figure 17 shows deposit latencies across the whole spread of disciplines.

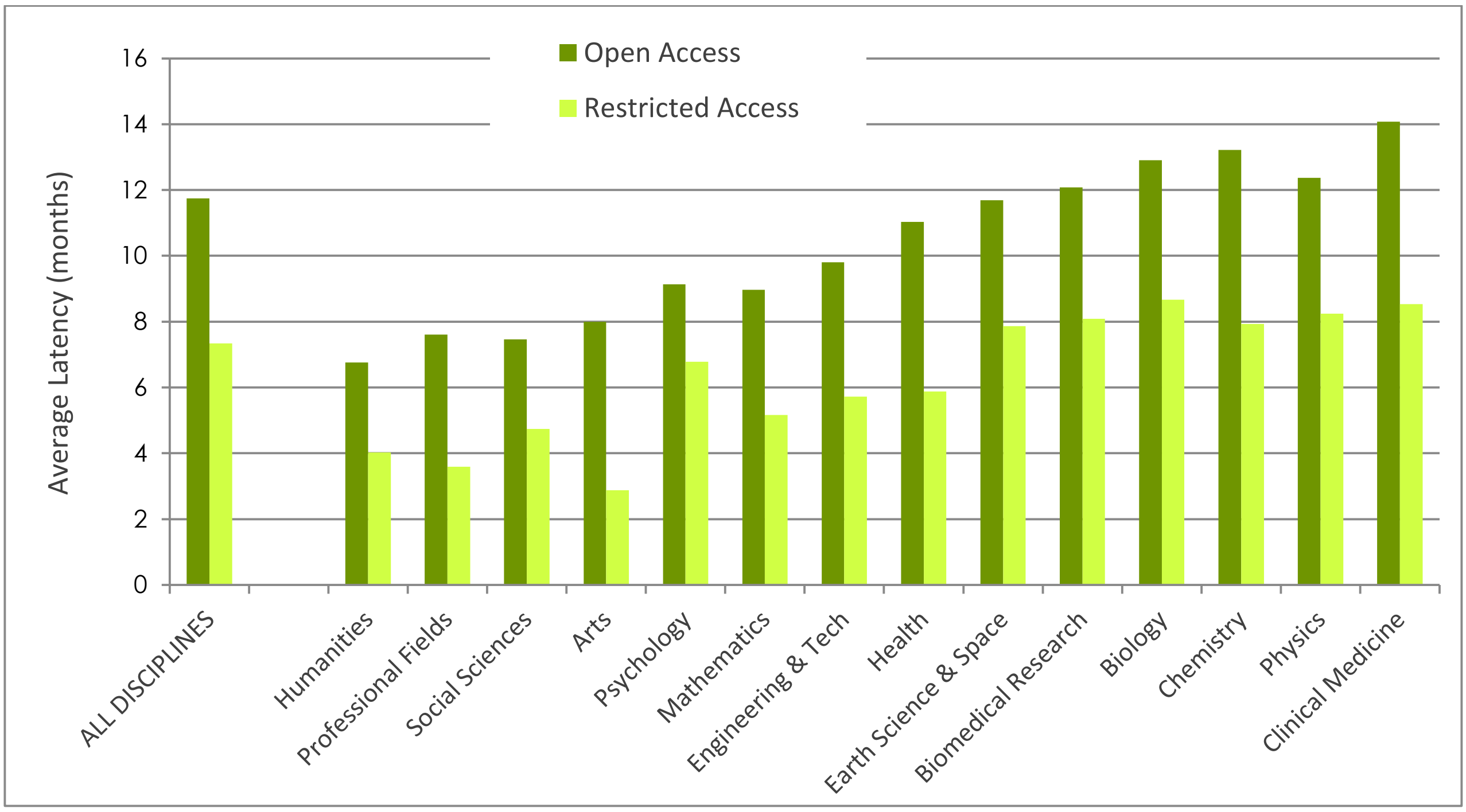

***Figure 17: Average deposit latencies by discipline for Open Access and Restricted Access deposits ordered by average Full-Text latency***

### *5.3.2 Deposit latency in mandated and non-mandated institutions*

Figure 18 below shows the data for deposit latencies in mandated and non-mandated institutions. Open Access items tend to be deposited later than Restricted Access items (have longer latencies) and latency periods tend to be longer in mandated institutions than in non-mandated ones (but deposit rates themselves are four times higher when deposit is mandated). Restricted Access deposits also tend to be converted to Open Access deposits after a delay, most likely because such deposits are initially set as Restricted Access because of publisher embargoes.

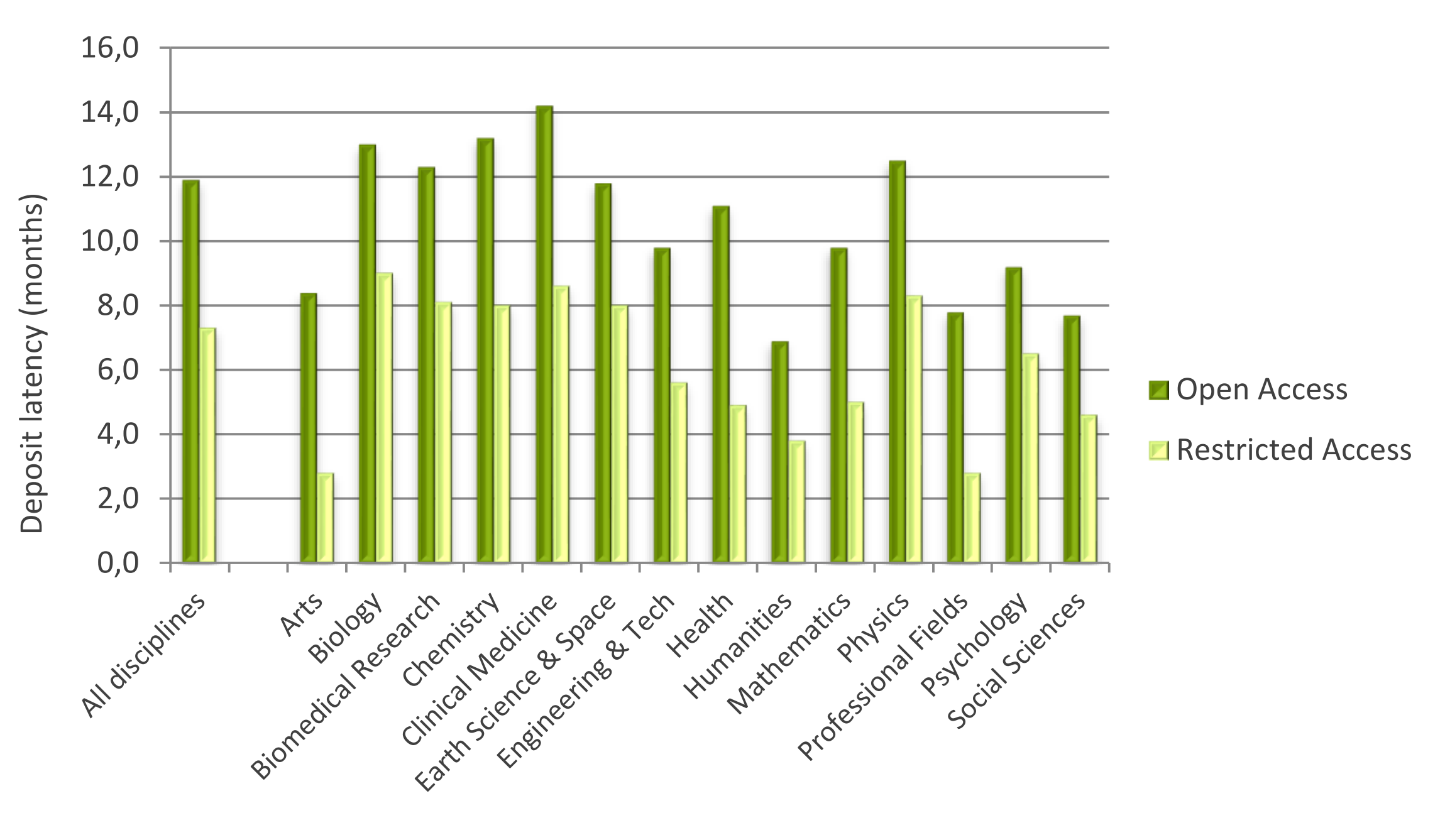

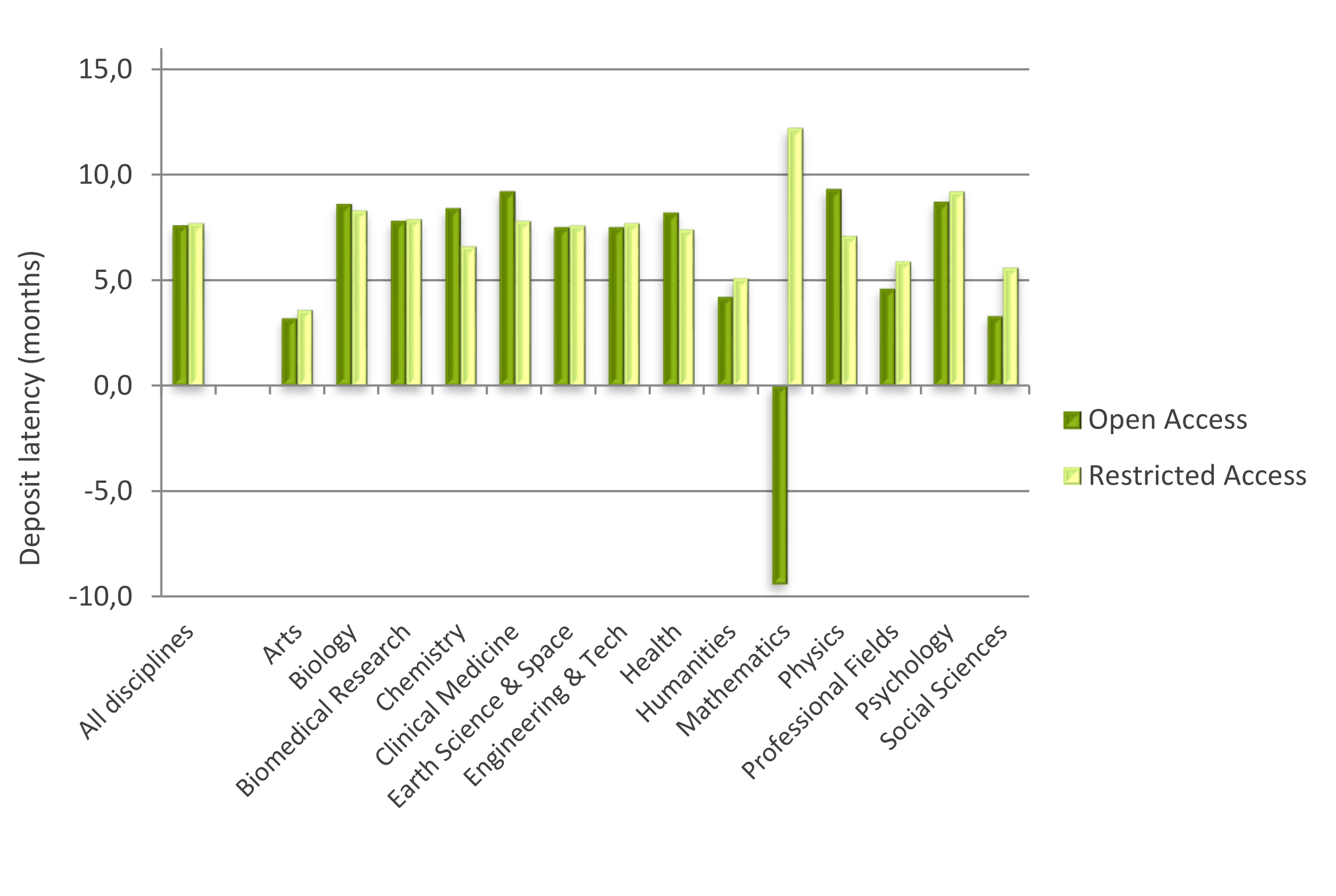

***Figure 18: Deposit latencies for mandated (top) and non-mandated (bottom) institutions, by discipline***

The deposit latency is shorter for non-mandated institutions than for mandated ones. This is likely to be because authors who deposit spontaneously are self-motivated to do it as early as possible, while those who deposit because it is mandated see no reason to do so early. Mathematics is a clear example: the early depositors are doing it because *they want to*, and they do it well before publication. The mandated depositors do it later. However, non-mandatory deposits number only a quarter of mandatory ones.

### *5.3.3 Deposit latencies for individual institutions*

Table 9 shows the deposit latencies for the top 20 institutions with the shortest latencies, mandated or non-mandated.

| Institution | Country | Number of articles in WoK | Total FT (OA + RA) | Open Access | Restricted Access |
|---|---|---|---|---|---|
| Bucknell University | USA | 349 | **0.5** | 0.5 | |
| Queen Margaret University, Edinburgh | UK | 150 | **2.1** | 1.6 | 2.2 |
| University of Southampton | UK | 7,916 | **3.0** | 3.0 | 3 |
| Katholieke Universiteit Leuven | Belgium | 11,293 | **3.3** | 2.6 | 3.9 |
| University of Bath | UK | 2847 | **3.5** | 3.6 | 2 |
| Imperial College London | UK | 15,462 | **3.7** | 3.7 | |
| University of Lincoln | UK | 1,536 | **3.8** | 2.7 | 4.1 |
| University of Edinburgh | UK | 10,201 | **4.0** | 4.0 | |
| University of Strathclyde | UK | 3,237 | **4.1** | 4.1 | 4.3 |
| Queensland University of Technology | Australia | 3,558 | **4.6** | 4.6 | 4.4 |
| National Institute of Oceanography | India | 462 | **4.8** | 4.8 | |
| University of Liege | Belgium | 4240 | **4.9** | 5.1 | 4.7 |
| Duke University | USA | 14,773 | **5.0** | 5.0 | |
| University of Warwick | UK | 5,464 | **5.2** | 5.3 | 2.2 |
| Birkbeck College, University of London | UK | 501 | **6.2** | 5.6 | 8.3 |
| University of Salford | UK | 822 | **7.0** | 5.8 | 7.4 |
| University of Abertay Dundee | UK | 2,740 | **7.2** | 7.2 | |
| Malmö University | Sweden | 537 | **7.3** | 7.3 | |
| University of Helsinki | Finland | 10,655 | **7.7** | 7.7 | |
| Universidade do Minho | Portugal | 3,021 | **8.1** | 8.3 | 7.8 |
| **All mandated institutions** | | | **11.1** | **11.9** | |
| **Non-mandated institutions (sample)** | | | **7.6** | **7.4** | **7.3** |

*Table 9: Average deposit latencies in months for the 20 institutions with the shortest latencies*

Section 5.4 presents data on how specific criteria in a policy influence depositing behaviour, including deposit latency.

### *5.3.4 Deposit latency in terms of policy-related time periods*

Figure 19 shows – for all the deposited articles only – the proportion of all deposited articles that were deposited within each of these five policy-related time periods:

- Before publication date
- Within 6 months of publication date
- Between 6 and 12 months of publication date
- Between 12 and 24 months of publication date
- After 24 months of publication date

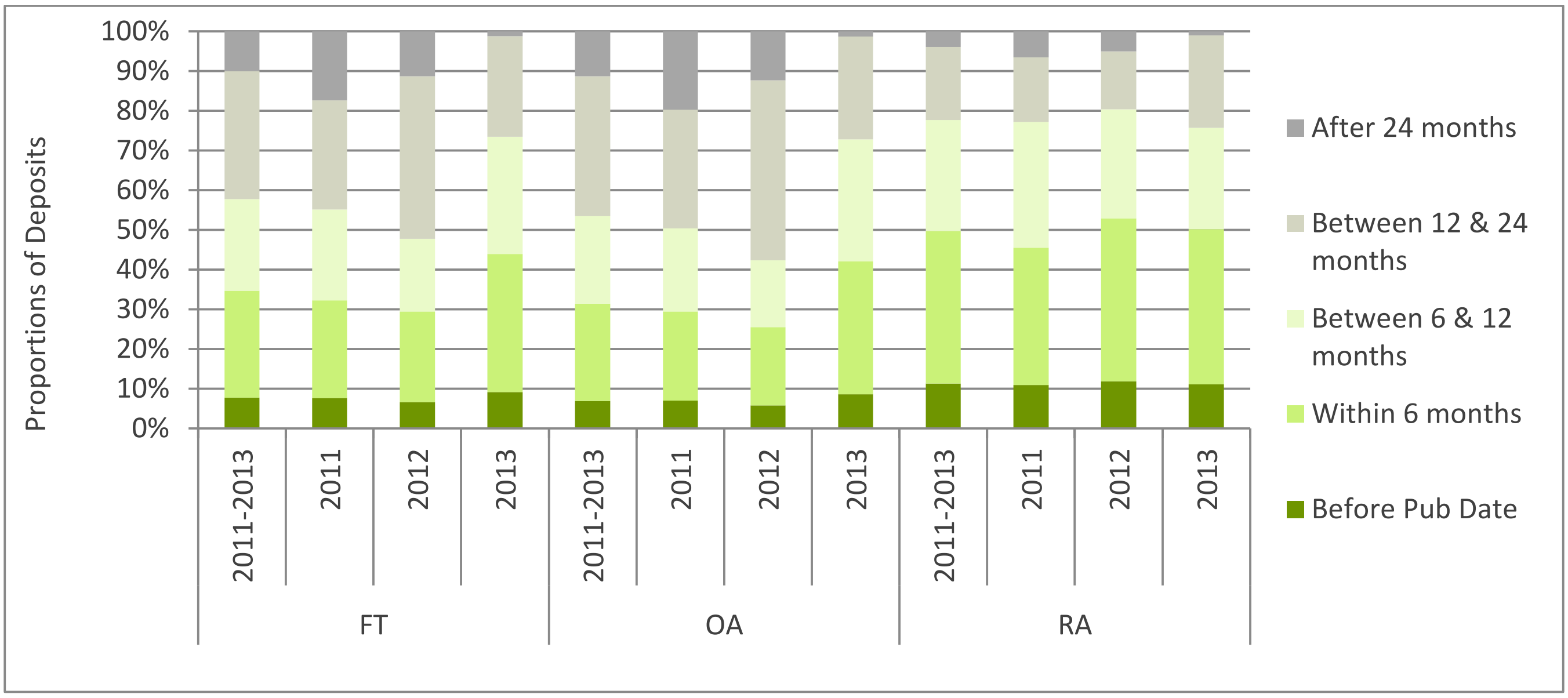

***Figure 19: Time periods in which deposits are made***

The most notable difference between 2011 and 2012 is the increase of the proportion of the Open Access deposits that were made between 12 and 24 months and the decrease of the proportion made after 24 months.

Restricted Access articles are deposited earlier than OA articles. Some of the later Open Access deposits are probably converted from Restricted Access: these were not double-counts, but the transitions were not tracked explicitly.

The data for 2013 were included in this figure but cannot properly be compared with 2011 and 2012 because the time window across which the average is calculated was shortest for 2013 (the data were collected in the third quarter of 2014). To make the three years comparable, we can take into account only the proportions of the deposited articles that were deposited up to 1 year after publication. These data are shown in Figure 20. Here the pattern appears to be unchanged across the 3 years, for OA as well as for RA.

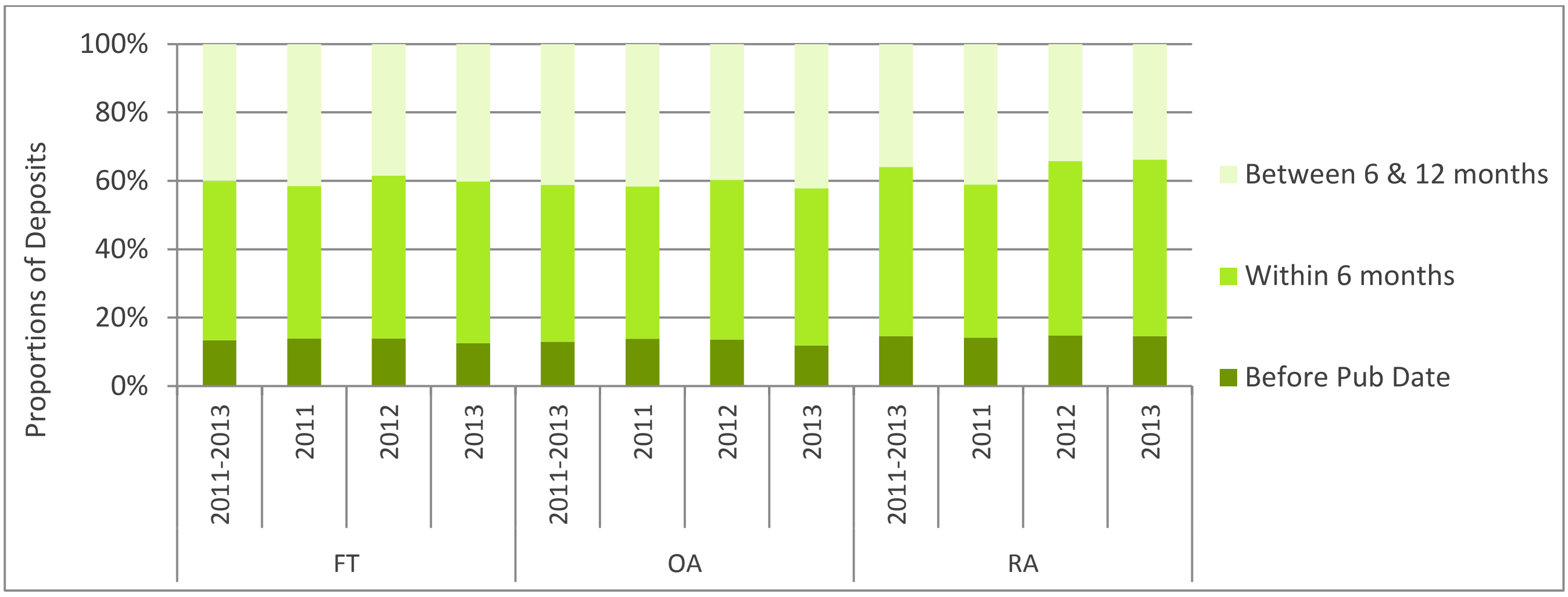

*Figure 20: Time periods within which deposits are made within one year after publication*

### 5.3.5 First year latency scores

Average deposit latency is a misleading way of comparing the three publication years. The latency for 2013 could been shorter, not because average latency is getting shorter from year to year, but just because the 2013 time window for calculating the average latency, which is only 2013-2014, is shorter than the window for 2011, which is 2011-2014. We accordingly computed a score based on only the deposits made within one year of publication for each year. This score is the sum of the weighted proportion in each of the 3 intervals of time (before publication, within 6 months of publication and between 6 and 12 months) for all articles deposited. The weights assigned are respectively 1, 2/3 and 1/3. Deposits done after 12 months are not considered. This score makes the 3 years of publication from 2011-2013 more comparable. The only detectable change seems to be that the RA deposits may be occurring a little earlier.

## 5.4 Effectiveness of particular policy conditions

To assess the effectiveness of policy conditions, we carried out a multiple regression analysis to test the correlation of policy conditions (independent variables) with deposit percentage and deposit latency (dependent variables). The full methodological approach is given in Appendix 2.

### 5.4.1 Deposit rate in relation to policy criteria

Our classification scheme for OA policies has 13 *conditions*, each of which has multiple (from 2 to 6) *values* (‘options’). Because not all the options have the same importance, we first gave *a priori* weights to each of the options: these weights are shown in Table 10. We then computed the pairwise Pearson correlation between each of the thirteen conditions and the deposit rates for Open Access, Restricted Access and Full-Text (OA + RA) items.

| Independent variables | Conditions | Condition options | Number of policies | Option Weight I | Option Weight II |
|---|---|---|---|---|---|
| Research evaluation | Is deposit a precondition for research evaluation (the ''Liège/HEFCE Model')? | Yes | 6 | 100% | 100% |
| | | Not specified | 93 | 0% | 0% |
| | | No | 32 | 0% | 0% |
| Must deposit | Deposit of item | Required | 93 | 100% | 100% |
| | | Requested | 26 | 10% | 0% |
| | | Not specified | 12 | 0% | 0% |
| Must make OA | Making deposited item Open Access | Required | 59 | 100% | 100% |
| | | Requested or recommended | 34 | 10% | 0% |
| | | Not mentioned | 28 | 0% | 0% |
| | | Other | 10 | 0% | 0% |
| Can not waive deposit | Can deposit of item be waived? | No | 28 | 100% | 100% |
| | | Not specified | 56 | 10% | 0% |
| | | Yes | 20 | 0% | 0% |
| | | Not applicable | 27 | 50% | 0% |
| Can not waive OA | Can making the deposited item OA be waived? | No | 15 | 100% | 100% |
| | | Not specified | 79 | 10% | 0% |
| | | Yes | 37 | 0% | 0% |
| Can not waive rights retention | Can author waive giving permission to make the article Open Access (where policy is based on faculty giving institution the right to make item OA) | Not applicable | 56 | 100% | 100% |
| | | No | 35 | 100% | 100% |
| | | Yes | 28 | 0% | 0% |
| | | Not specified | 12 | 10% | 0% |
| Deposit immediately | Date of deposit | No later than time of acceptance | 14 | 100% | 100% |
| | | No later than publication date | 17 | 20% | 0% |
| | | By end of the policy-specified embargo | 6 | 10% | 0% |
| | | When publisher permits | 5 | 5% | 0% |
| | | Not specified | 82 | 0% | 0% |
| | | Other | 7 | 0% | 0% |
| Make OA immediately | Date deposit to be made Open Access | Acceptance date | 4 | 100% | 100% |
| | | Publication date | 3 | 75% | 100% |
| | | By end of policy-permitted | 14 | 50% | 0% |
| | | embargo | | | |
| | | As soon as deposit is completed | 2 | 5% | 0% |
| | | When publisher permits | 33 | 5% | 0% |
| | | Not mentioned | 72 | 0% | 0% |
| | | Other | 3 | 0% | 0% |
| Embargo permitted: STEM | Policy's permitted embargo length for Science, Technology, Engineering and Mathematics | Not specified | 116 | 100% | 100% |
| | | 0 months | 1 | 100% | 100% |
| | | 6 months | 7 | 50% | 50% |
| | | 12 months | 6 | 5% | 5% |
| | | Longer | 1 | 0% | 0% |
| Embargo permitted: HaSS | Policy's permitted embargo length for Humanities and Social Sciences | Not specified | 117 | 100% | 100% |
| | | 0 months | 1 | 100% | 100% |
| | | 6 months | 6 | 50% | 50% |
| | | 12 months | 7 | 50% | 50% |
| Deposit in institutional repository | Locus of deposit | Institutional repository | 123 | 100% | 100% |
| | | Any suitable repository | 2 | 0% | 0% |
| | | Not specified | 6 | 0% | 0% |
| Must retain rights | Rights holding | Author retains key rights | 38 | 100% | 100% |
| | | Author grants key rights to institution | 1 | 100% | 100% |
| | | Institution or funder retains key rights | 1 | 100% | 100% |
| | | None of these | 37 | 0% | 0% |
| | | Not mentioned | 54 | 0% | 0% |
| Open licensing conditions | Open licensing conditions | Does not require any re-use licence | 72 | 100% | 100% |
| | | Other | 24 | 50% | 50% |
| | | Not specified | 11 | 50% | 50% |
| | | Requires CC-BY or equivalent | 1 | 0% | 0% |
| | | Requires CC-BY-NC or equivalent | 2 | 0% | 0% |
| | | Requires an open licence without specifying which one | 21 | 0% | 0% |

***Table 10: The 13 Open Access policy conditions and weights assigned to the options under each condition***

We eliminated two of the conditions (*Embargo permitted: STEM* and *Embargo permitted: HaSS*) because so few policies mentioned them. The pairwise correlations of eleven of the thirteen conditions are shown in Table 11.

| Variables | Full-text deposit rate | | Open Access deposit rate | | Restricted Access deposit rate | |
|---|---|---|---|---|---|---|
| | p | r | p | r | p | r |
| **Cannot waive deposit** | 0.020 | 0.244 | 0.209 | 0.047 | 0.472 | 0.119 |
| **Research evaluation** | 0.025 | 0.235 | 0.292 | 0.112 | 0.312 | 0.166 |
| **Cannot waive rights retention** | 0.154 | 0.151 | 0.153 | 0.155 | 0.484 | 0.115 |
| **Must make OA** | 0.312 | 0.067 | 0.093 | 0.177 | 0.385 | -0.143 |
| **Must deposit** | 0.122 | 0.163 | 0.047 | 0.209 | 0.648 | 0.075 |
| **Cannot waive OA** | 0.490 | -0.073 | 0.451 | -0.038 | 0.641 | -0.077 |
| Deposit immediately | 0.658 | 0.079 | 0.652 | 0.048 | 0.488 | 0.114 |
| Make OA Immediately | 0.676 | -0.044 | 0.622 | -0.052 | 0.743 | -0.054 |
| Must retain rights | 0.709 | 0.040 | 0.558 | 0.062 | 0.646 | 0.076 |
| Mandate Age | 0.592 | 0.058 | 0.994 | -0.001 | 0.488 | 0.118 |
| Open licensing | 0.516 | 0.098 | 0.241 | 0.124 | 0.959 | -0.009 |

***Table 11: pairwise correlations for eleven OA policy conditions (based on 'Option Weight I') with Open Access, Restricted Access and Full-text deposit rate***

r: Pearson's correlation coefficient (between -1 and 1)
p: Probability of observed correlation by chance when the real population correlation is zero
**Bold**: Potential independent variables retained for multiple regression analysis (see below)

For the next regression analysis we further reduced the remaining eleven conditions by eliminating the five with correlation coefficients between -0.1 and 0.1 (*Deposit immediately, Make OA immediately, Must retain rights, Mandate age* and *Open licensing*). That left only the six conditions indicated in bold in Table 11):

- Cannot waive deposit
- Research evaluation
- Cannot waive rights retention
- Must make OA
- Must deposit
- Cannot waive OA

To further increase predictive power, we updated the initial weights of the options for this second regression analysis. For the first analysis we had initialized the weights according to the column headed 'Option Weight I' in Table 8. In the update we transformed the options into dichotomous (all-or-none) ones – either 100% or 0%. We assigned 100% to those we hypothesised to be stronger options (e.g., if they required authors to deposit articles and/or to deposit them immediately) and 0% to weaker ones. These are shown in the column headed 'Option Weight II' in Table 10.

This transformation reduces the number of options from up to 6 to 2 for each condition. Although this does lose some information about intermediate options (where policy strength is probably somewhere between 0% and 100%) the advantage it brings is that there is a higher number of instances for each of the remaining two options. Collapsing the options should have little effect on the pairwise correlations.

We then carried out multiple regression analyses using ~~the~~ Negative Binomial Regression (NBR) ~~model~~ to test how well the remaining six policy conditions (independent variables) correlate with the three deposit rate measures: Open Access, Restricted Access and Full-Text deposits (dependent variables). The results are shown in Table 12.

In this table, *E(β)* is the incidence rate ratio, that is the degree of increase in the dependent variable for a 1-unit increase in the predictor variable. The correlation is positive when *E(β)* is greater than 1 and negative when *E(β)* is between 0 and 1. Table 12 also shows the pairwise correlations for each of the six policy conditions (using 'Option Weight II') with Open Access, Restricted Access and Full-Text (OA + RA) deposit rate.

| Dependent Variable | Independent Variables | NBR (multiple regression) | | Pearson Correlation (pairwise correlation) | |
|---|---|---|---|---|---|
| | | p | Exp(β) | p | r |
| Full-Text deposit rate | Cannot waive deposit | **0.026** | 1.932 | **0.002** | 0.324 |
| | Research evaluation | 0.356 | 1.570 | **0.025** | 0.235 |
| | Cannot waive rights retention | 0.266 | 1.298 | 0.159 | 0.149 |
| | Must deposit | 0.547 | 1.166 | 0.122 | 0.163 |
| | Must make Open Access | Near zero | Near zero | Near zero | Near zero |
| | Cannot waive Open Access | 0.073 | 0.514 | 0.469 | -0.077 |
| Open Access deposit rate | Cannot waive deposit | 0.089 | 1.647 | **0.006** | 0.287 |
| | Research evaluation | 0.570 | 1.320 | 0.292 | 0.112 |
| | Cannot waive rights retention | 0.492 | 1.185 | 0.149 | 0.153 |
| | Must deposit | 0.307 | 1.324 | **0.047** | 0.209 |
| | Must make Open Access | 0.466 | 1.218 | 0.107 | 0.170 |
| | Cannot waive Open Access | 0.146 | 0.573 | 0.366 | -0.032 |
| Restricted Access deposit rate | Cannot waive deposit | 0.166 | 1.895 | 0.266 | 0.183 |
| | Research evaluation | Near zero | Near zero | Near zero | Near zero |
| | Cannot waive rights retention | Near zero | Near zero | Near zero | Near zero |
| | Must deposit | 0.743 | 1.174 | 0.648 | 0.075 |
| | Must make Open Access | 0.246 | 0.584 | 0.350 | -0.154 |
| | Cannot waive Open Access | 0.263 | 0.418 | 0.518 | -0.107 |

***Table 12: Negative Binomial Regression and Pairwise Correlations for six Open Access policy conditions (based on Option weight II) with the deposit rates for Open Access, Restricted Access and Full-Text items (OA + RA) deposit***

**r**: Correlation coefficient (between -1 and 1)

p: Probability of observed correlation by chance when the real population correlation is zero

**Black**: positive correlation **Red**: negative correlation **Blue**: Significant correlation ($p < 0.05$)

**E(β)**: the incidence rate ratio is rate of increase in the dependent variable, for a 1 unit increase in the predictor variable. The correlation is positive when E(β) is greater than 1 and negative when E(β) is between 0 and 1.

The graphs below that comprise Figure 21 show all the correlations in the form of the average rates for Open Access, Restricted Access and Full-Text (OA + RA) items for each policy condition option, for the years 2011-2013.

The important findings to note are the trends. These are, in general, the same for each policy condition, and where there are near-zero correlations there are simple explanations for them (see Section 5.4.3).

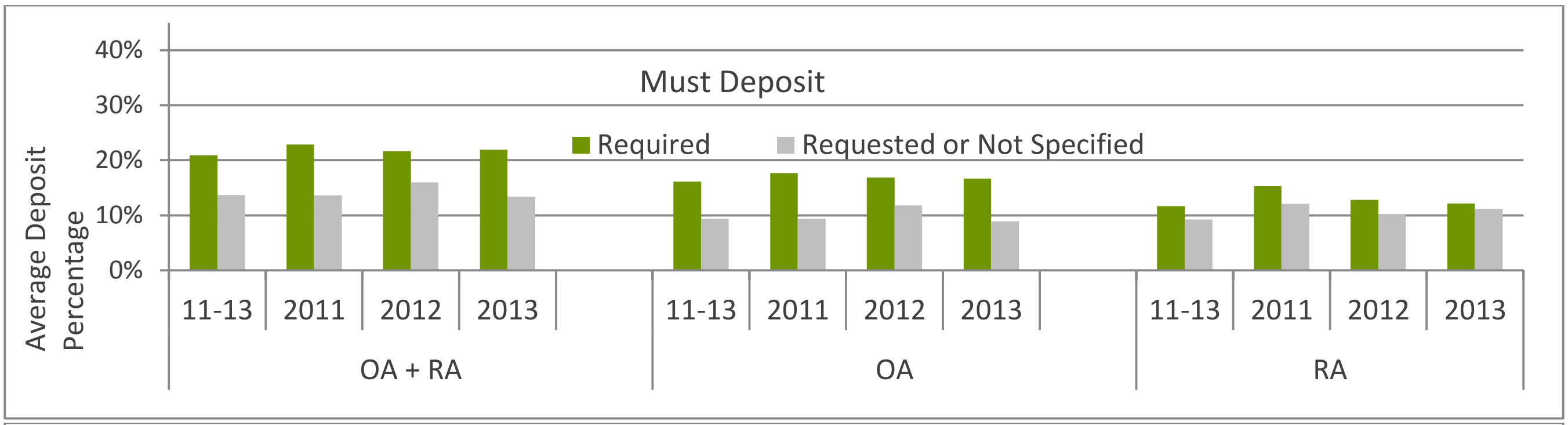

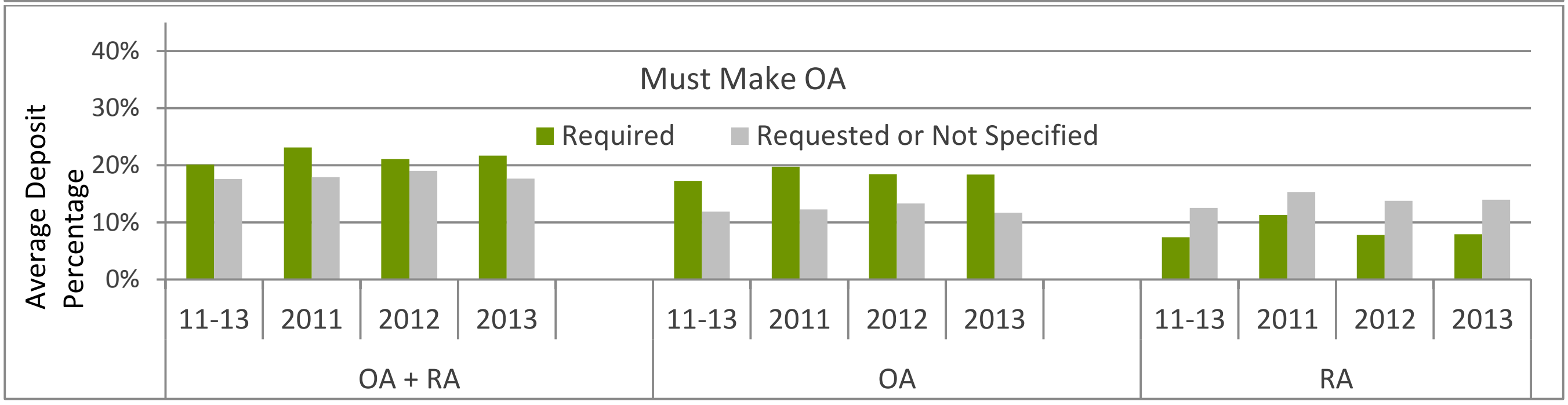

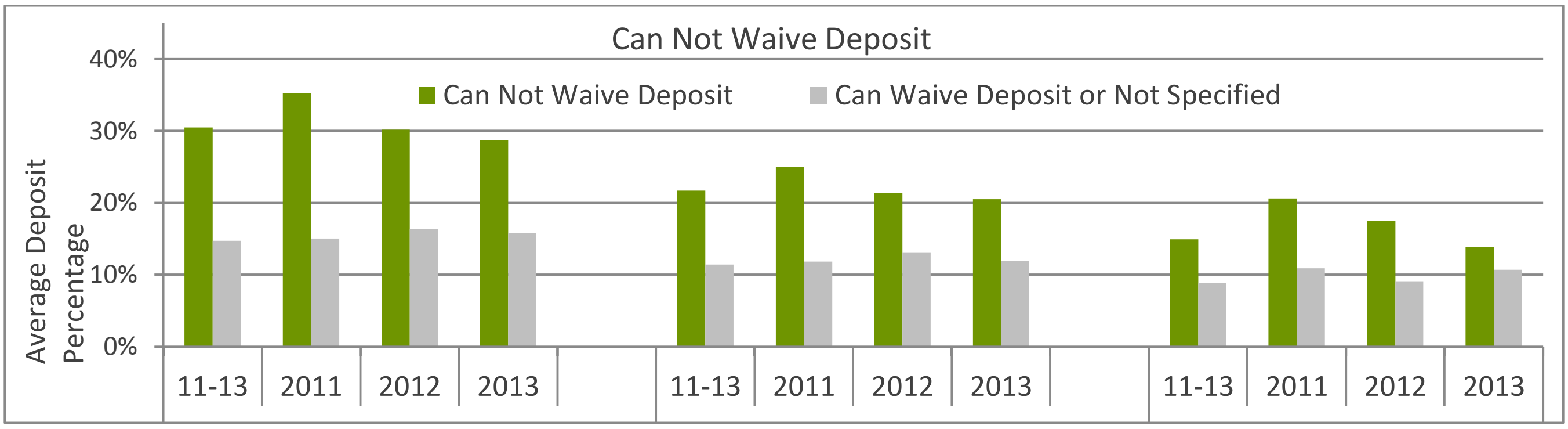

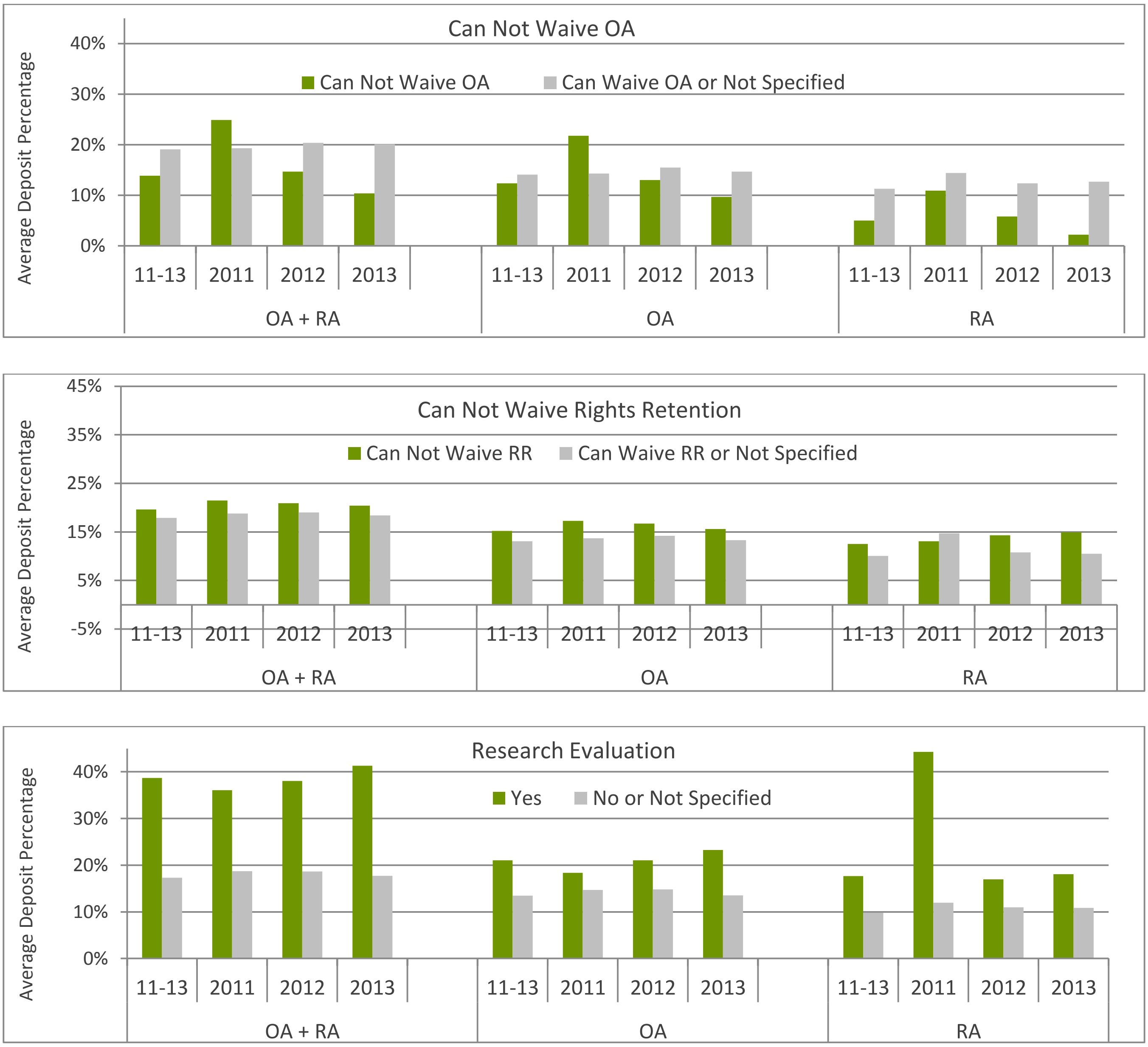

***Figure 21: Average percentage deposit by policy condition options***

### *5.4.2 Deposit latency in relation to policy criteria*

The same analysis was carried out using the 13 independent variables to test correlation with First Year Latency Score (Y1 Latency Score). To make this intuitive, the latency scores are coded so that the higher the score the earlier (better) the deposit.

We carried out pairwise correlations to test the potential associations between each of the 11 conditions (based on *a priori* weights 'Option Weight I: see Table 10) and the Y1 Latency Score, for Full-Text, Open Access and Restricted Access deposits. The results are shown in Table 13.

| Independent variables | Full-text First Year Latency Score | | Open Access First Year Latency Score | | Restricted Access First Year Latency Score | |
|---|---|---|---|---|---|---|
| | p | r | p | r | p | r |
| **Mandate age** | 0.005 | 0.306 | 0.007 | 0.295 | 0.154 | 0.250 |
| **Cannot waive rights retention** | 0.147 | 0.159 | 0.099 | 0.180 | 0.766 | 0.051 |
| **Deposit immediately** | 0.174 | 0.149 | 0.218 | 0.135 | 0.479 | 0.122 |
| Can not waive deposit | 0.927 | 0.010 | 0.898 | 0.014 | 0.691 | 0.069 |
| Research evaluation | 0.501 | 0.074 | 0.556 | 0.065 | 0.357 | 0.158 |
| Must make OA | 0.754 | 0.035 | 0.913 | 0.012 | 0.763 | 0.052 |
| Must deposit | 0.557 | 0.065 | 0.748 | 0.035 | 0.610 | 0.088 |
| Make OA Immediately | 0.748 | 0.035 | 0.893 | 0.015 | 0.969 | -0.007 |
| Must retain rights | 0.724 | -0.039 | 0.853 | -0.020 | 0.299 | -0.178 |
| Can not waive OA | 0.809 | -0.027 | 0.748 | -0.035 | 0.496 | -0.117 |
| Open licensing | 0.887 | -0.016 | 0.714 | -0.040 | 0.132 | 0.256 |

***Table 13: Pairwise Correlations between Eleven OA Policy Conditions (based on "Option Weight V1") and First Year Latency Score (OA, RA and FT)***

r: Correlation coefficient (between -1 and 1)
p: Probability of error
**Bold: Potential independent variables retained for multiple regression analysis**

Only the correlation between Mandate Age and latency is significant. However, two other independent variables (*Can not waive rights retention* and *Deposit immediately*) look as if their correlations with latency might become significant if the sample size were bigger. The eight other variables with a correlation coefficient under 0.1 were excluded from multiple regression tests.

As before, we then carried out Negative Binomial Regression to test whether the three potential policy conditions (independent variables) jointly correlate with the three First Year Latency Scores (dependent variables) for Open Access, Restricted Access and Full-Text deposits. Table 14 shows the results of these NBR tests as well as the separate pairwise correlations between the three OA policy conditions (based on 'Option Weight 2') and the First Year Latency Scores (OA, RA and FT).

| Dependent Variable | Independent Variables | NBR (multiple regression) | | Pearson Correlation (pairwise correlation) | |
|---|---|---|---|---|---|
| | | p | Exp(β) | p | r |
| Full-Text First Year Latency Score | Mandate age | 0.181 | 1.075 | **0.005** | 0.306 |
| | Can not waive rights retention | 0.577 | 1.145 | 0.141 | 0.161 |
| | Deposit immediately | 0.623 | 1.186 | 0.134 | 0.164 |
| Open Access First Year Latency Score | Mandate age | 0.200 | 1.072 | **0.007** | 0.295 |
| | Can not waive rights retention | 0.489 | 1.183 | 0.096 | 0.182 |
| | Deposit immediately | 0.662 | 1.163 | 0.159 | 0.154 |
| Restricted Access First Year Latency Score | Mandate age | 0.413 | 1.057 | 0.154 | 0.250 |
| | Can not waive rights retention | 0.819 | 1.087 | 0.735 | 0.058 |
| | Deposit immediately | Near zero | Near zero | 0.570 | 0.098 |

***Table 14: Negative Binomial Regression and Pairwise Correlations for Open Access policy conditions (based on Option Weight II) with First Year Latency Score for Open Access, Restricted Access and Full-Text items***

**r**: Pearson's correlation coefficient (between -1 and 1)
p: Probability of error
**Bold**: significant correlation ($p < 0.05$)
**E(β)**: the incidence rate ratio is rate of increase in the dependent variable, for a 1 unit increase in the predictor variable. The correlation is positive when E(β) is greater than 1 and negative when E(β) is between 0 and 1.

There are not yet enough OA policies to design a model that can predict latency with credible probability. However, three promising trends are worth pointing out. NBR shows that none of the three correlations with latency was individually significant, but including all three independent variables in the regression contributes to a better fit of the model without increasing the error p value of the overall model.

- ***Mandate age***: authors in institutions with older OA policies are more likely to deposit their articles earlier. This could be explained by the fact that recent policies have had less time (and less chance) to be complied with and made habitual. Also, for more recent policies, when a mandate is first adopted, many authors deposit not only their current articles but, at the same time, also their backlog of older ones. This would also increase latency for the publications from the earlier years.
- ***Cannot waive rights retention***: When authors cannot waive giving permission to make their article OA, they are more likely to deposit their articles earlier (as OA and also as RA).
- ***Deposit immediately****:* There is no correlation between *Deposit immediately* and RA latency, probably because RA deposits, even if they are made early, are converted to OA deposits after the end of the publisher embargo. However, when deposit is required no later than the time of acceptance, authors deposit their articles as OA earlier than for all other options related to date of deposit.

**Figure 21** shows averages for *First Year Latency Score* (OA, RA and FT) for each policy condition option related to *Cannot waive rights retention* and *Deposit immediately*.

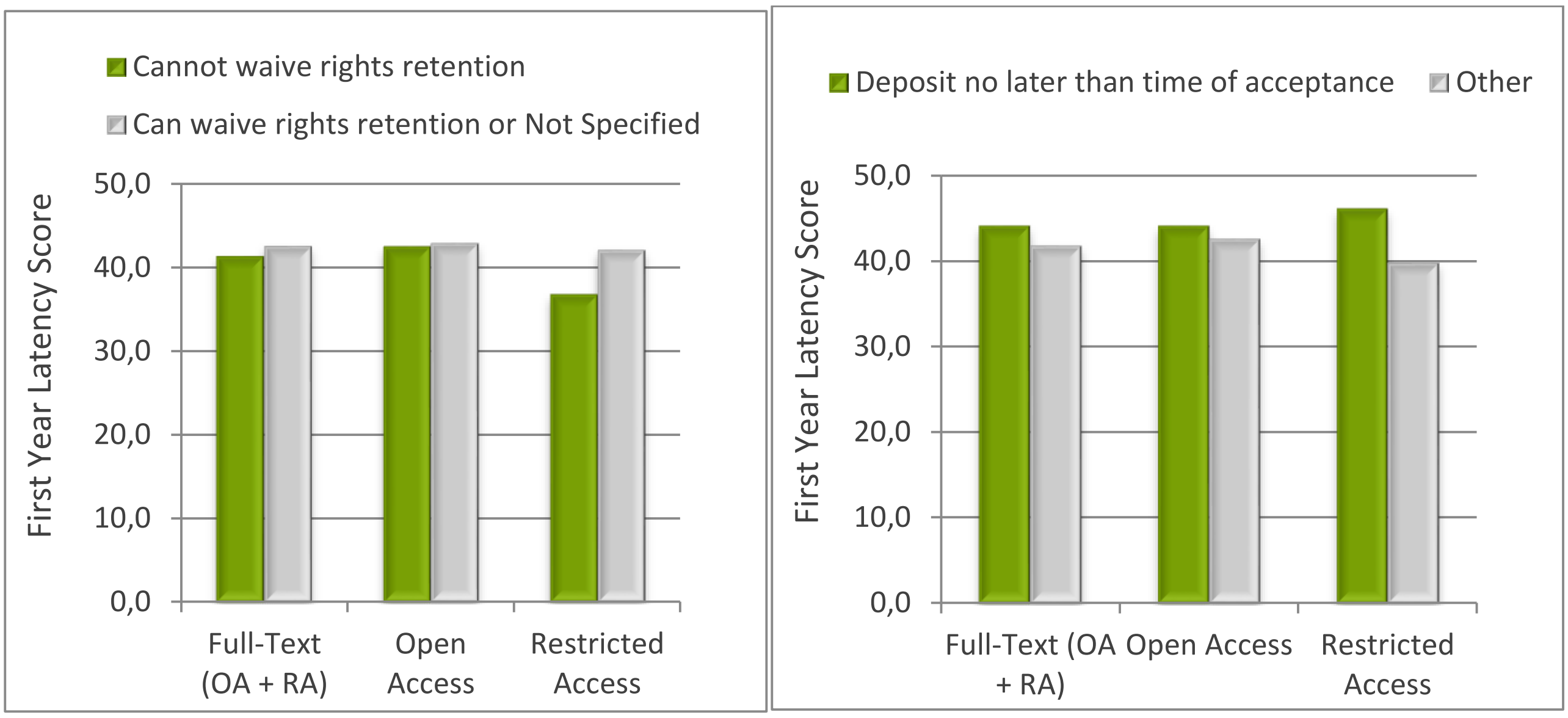

***Figure 21: First Year latency Score by policy condition options Cannot waive rights retention and Deposit immediately (2001-2013)***

### *5.4.3 Summary of the results of the regression analyses*

Although the number of OA policies is not yet large enough to design a full jointly predictive model that reaches statistical significance, some correlation trends are already apparent in the data.

The regression analyses demonstrate the following:

| | Correlated with | Regression | Significant correlation? | Comments |
|---|---|---|---|---|
| Must deposit | Open Access deposit rate<br>Restricted Access deposit rate | Pairwise<br>NBR, pairwise | Yes | As expected, when deposit is mandated, the deposit rate for Open Access items is higher. |
| Cannot waive deposit | Full-Text deposit rate<br>Open Access deposit rate<br>Restricted Access deposit rate | NBR, pairwise<br>Pairwise<br>NBR, pairwise | Yes (NBR and pairwise)<br>Yes | When authors cannot waive deposit the deposit rate is likely to be higher for Full-Text and Open Access deposits. |
| Research evaluation | Full-Text deposit rate | Pairwise | Yes | When deposit is required for research evaluation, authors are more likely to deposit their full-text as Open Access articles |
| Must make OA | Open Access deposit rate<br>Restricted Access deposit rate | NBR, pairwise<br>NBR, pairwise | <br>Negative correlation | When OA deposit is required rather than requested, OA deposit is more likely. However, this condition also makes authors less likely to deposit their article as Restricted Access. Authors probably wait until an embargo has elapsed, and then make the deposit Open Access directly. |
| Cannot waive OA | Full-Text deposit rate<br>OA deposit rate<br>Restricted Access deposit rate | NBR, pairwise<br>NBR, pairwise<br>NBR, pairwise | Negative correlation<br>Negative correlation<br>Negative correlation | When authors are required to make their deposits Open Access rather than Restricted Access the deposit rate is lower. This is probably because authors are reluctant to ignore a publisher OA embargo. |
| Cannot waive rights retention | Full-Text deposit rate<br>OA deposit rate | NBR, pairwise<br>NBR, pairwise | | There is a small but non-significant increase in Open Access deposit rate if authors are required to retain the rights they need for Open Access and are not permitted to waive this (this applies in 8 cases at the time of writing). Retention of the necessary rights means authors are confident that they are contractually entitled to deposit their articles and make them Open Access |

***Table 15: Summary of the correlations between deposit and policy condition options***

*Black = positive correlation* *Red = negative correlation*

As the numbers stand at the moment (March 2015), there are not yet enough OA policies to test whether other policy conditions would further contribute to mandate effectiveness. The current findings, however, already suggest that it would be useful for future mandates to adopt these conditions so as to maximise the growth of OA.

Moreover, this analysis provides a list of criteria around which policies should align:

- Must deposit (i.e. deposit is mandatory)
- Deposit cannot be waived
- Link deposit with research evaluation

## Appendix 1

The following criteria were developed to classify policies in ROARMAP.

**Institutional particulars**

| | |
|---|---|
| *Location* | - Continent/region<br>- Country |
| *Policymaker type* | - Funder<br>- Research organisation (e.g. university or research institution)<br>- Funder and research organisation<br>- Multiple research organisations<br>- Sub-unit of research organisation (e.g. department, faculty or school)<br>- Unspecified |
| *Policymaker name* | - [free text] |
| *Policymaker URL* | - [free text] |
| *Policy URL* | - [free text] |
| *Repository URL* | - [free text] |
| *Source of policy* | - Administrative or management decision<br>- Faculty vote<br>- Not mentioned<br>- Other |

**Dates relating to policy implementation**

| | |
|---|---|
| *Policy adoption date* | - [free text] |
| *Policy effective date* | - [free text] |
| *Last revision date* | - [free text] |

**Criteria for deposit and licensing conditions**

*Deposit of item*
- Required
- Requested
- Unspecified

*Locus of Deposit*
- Institutional Repository
- Subject repository
- Any suitable repository
- Not Specified

*Date of deposit*
- No later than the time of acceptance
- No later than the publication date
- By end of policy-specified embargo
- When publisher permits
- Not Specified
- Other

*Option to waive deposit*
- Yes
- No
- Not specified
- Not Applicable

*Requirement to make item Open Access*
- Required
- Requested or recommended
- Not Mentioned
- Other
- Not specified

*Option to waive Open Access on deposited item*
- Yes
- No
- Not specified
- Not Applicable

| | |
|---|---|
| *Date deposit should be made Open Access* | |
| | - Not specified |
| | - Acceptance date |
| | - Publication date |
| | - By end of policy-permitted embargo |
| | - When publisher permits |
| | - As soon as the deposit is completed |
| | - Not Mentioned |
| | - Other |
| *Condition for research evaluation (the 'Liège/HEFCE Model')* | - Yes |
| | - No |
| | - Not specified |
| *Option for author to waive giving permission to make item Open Access* | - Yes |
| | - No |
| | - Not specified |
| | - Not Applicable |
| *Open licensing conditions* | - Does not require any re-use licence |
| | - Requires an open licence without specifying which one |
| | - Requires CC-BY or equivalent |
| | - Requires CC-BY-NC or equivalent |
| | - Requires a different open licence |
| | - Other |
| | - Not specified |

***Rights holding***

| | |
|---|---|
| *Rights holding* | - Author grants key rights to institution |
| | - Institution or funder retains key rights |
| | - Author retains key rights |
| | - None of these |

- Not Mentioned
- Not specified

*Can granting key rights be waived*

- Yes
- No
- Not specified
- Not Applicable

***Embargo lengths and publishing options***

*Policy's permitted embargo length for science, technology medicine (STEM)*

- 0 months
- 6 months
- 12 months
- 24 months
- Longer
- Not Specified

*Policy's permitted embargo length for humanities and social sciences (HaSS)*

- 0 months
- 6 months
- 12 months
- 24 months
- Longer
- Not Specified

*Can maximum allowable embargo length be waived?*

- Yes
- No
- Not specified
- Not Applicable

*Gold OA publishing option*

- Required
- Recommended as an alternative to Green self-archiving
- Permitted alternative to Green self-archiving
- Not Specified
- Other

*Funding for APCs where charged by journals*

- Funder allows APCs to be paid from research grant
- Funder provides specific additional funding for APCs
- Institution provides funding
- Not Mentioned
- Other

*APC fund URL*
*(where available)* - [free text]

## Appendix 2: Policy effectiveness methodologies

The study on Open Access (OA) policy effectiveness was conducted in November 2014 and was based on institutional mandatory OA policies indexed by ROARMAP (the Registry of Open Access Repositories Mandatory Archiving Policies). Out of 244 institutional mandates, only the 122 that had been adopted by 2011 or earlier were analysed (because more recent mandates would not yet have been in effect within our 2011-2013 comparison window).

Our interest was in which policy parameters contributed significantly to mandate effectiveness, but to get some sense of how effective mandates as a whole are, compared to non-mandatory policies, we also analysed the 10 (out of 142) institutions with the highest research output between 2011 and 2013 that have a non-mandatory OA policy. (This subset is not representative of all non-mandatory policies and is only used for some specific comparisons.)

***Institutional Repository content and the percentage of Open Access, Restricted Access, Full-Text and Metadata-Only deposits***

The bibliographic metadata for all journal articles published between 2011 and 2013 by authors at these 132 institutions (N = 347,880) in journals indexed by Thompson-Reuters database (WoK) were extracted from the WoK database. Our robot harvested Institutional Repository (IR) contents to check *whether* and *when* the Full-Text (FT) for each of those WoK articles was deposited as either Open Access [OA] or Restricted Access [RA]; the robot also collected data on the cases where Metadata Only (MO: no FT) were deposited, or where nothing was deposited at all. The IR metadata were extracted from ROAR (the *Registry of Open Access Repositories*) as well as directly from each IR's own website.

***Deposit Latency***

Publication date was estimated based on the Altmetrics database. When these data were not available, we used the WoK publication date and subtracted 5.26 months, which is the average difference in dates between Altmetrics and WoK. By subtracting the date when articles were deposited (as indicated by the IR metadata) from the date they were published, we calculated the Average Deposit Latency in months.

***Regression analysis***

In order to assess the effectiveness of OA policies, we carried out multiple regression analyses to test which policy conditions (independent variables) correlate with deposit rate and deposit latency (dependent variables).

We excluded 26 institutions that had fewer than 50 publications during the 2011-2013. We also excluded 8 institutions where the locus of deposit was not explicitly specified as being the institutional repository. This yielded 98 institutional policies for analysis, including the 10 non-mandated institutions.